\documentclass[reprint, 11pt,superscriptaddress,preprintnumbers,nofootinbib,aps]{revtex4}

\usepackage{graphicx}
\usepackage{amsmath,amssymb,amsfonts,amsthm,stmaryrd,mathtools,bm,physics,tensor}
\usepackage{color}
\usepackage{tikz}
\allowdisplaybreaks[1]
\usepackage[normalem]{ulem}

\usepackage[bookmarks,linktocpage, colorlinks=true, plainpages = false, citecolor = treegreen,  linkcolor=darkblue, urlcolor = darkblue, filecolor = blue]{hyperref} 
\usepackage[bookmarks,linktocpage, colorlinks=true, plainpages = false, citecolor = treegreen,  linkcolor=darkblue, urlcolor = darkblue, filecolor = blue]{hyperref} 

\def\be#1\ee{\begin{align}#1\end{align}}

\def\ba{\begin{eqnarray}}
	\def\ea{\end{eqnarray}}
\def\nn{\nonumber}

\definecolor{darkblue}{rgb}{0., 0.4, 0.8}
\definecolor{cadmiumred}{rgb}{1., 0., 0.22}
\definecolor{treegreen}{rgb}{0., 0.7, 0.3}
\definecolor{orchid}{rgb}{0.7., 0., 0.5}

\begin{document}

\title{Regular black holes from pure gravity in four dimensions}

\author{Johanna Borissova}
\email{j.borissova@imperial.ac.uk}
\affiliation{Abdus Salam Centre for Theoretical Physics, Imperial College London, London SW7 2AZ, UK}
\author{Ra\'ul Carballo-Rubio}
\email{raul.carballorubio@iaa.csic.es}
\affiliation{Instituto de Astrof\'isica de Andaluc\'ia (IAA-CSIC), Glorieta de la Astronom\'ia, 18008 Granada, Spain}

\begin{abstract}
\bigskip
{\sc Abstract:} We derive static spherically symmetric regular black holes as vacuum solutions to purely gravitational theories in four dimensions.
To that end, we construct four-dimensional non-polynomial gravities starting from subclasses of two-dimensional Horndeski actions. By construction, these theories possess second-order equations of motion on spherically symmetric backgrounds. We show that a subset of these non-polynomial gravities, referred to as non-polynomial quasi-topological gravities, admit single-function static spherically symmetric solutions whereby the metric function is determined by an algebraic equation. Solutions to these theories include the Hayward regular black-hole spacetime, for which a corresponding gravitational action is stated explicitly. 
\end{abstract}

\maketitle
\tableofcontents

\section{Introduction}

Regular black holes 
and horizonless compact objects are tentative alternatives to the classical black holes in general relativity, which inevitably exhibit spacetime singularities~\cite{Penrose:1964wq,Hawking:1970zqf,Hawking:1973uf}. Prominent examples of regular black holes are the Bardeen~\cite{Bardeen:1968bh}, Dymnikova~\cite{Dymnikova:1992ux} and Hayward~\cite{Hayward:2005gi} spacetimes. For overviews and reviews on the current status of non-singular black holes and mimickers in classical and quantum gravity, see, e.g., \cite{Carballo-Rubio:2025fnc,Buoninfante:2024oxl,Bambi:2025wjx,Lan:2023cvz,Platania:2023srt,Buoninfante:2022ild}. 

While the regularization of classical singularities can be expected to be induced by quantum-gravitational effects, such as higher-derivative and non-local operators entering the gravitational effective action, obtaining regular black holes as fully non-linear solutions to gravitational actions in four dimensions has remained elusive for decades. It is well-known that spherically symmetric regular black-hole spacetimes can be regularized by coupling general relativity to non-linear electrodynamic sources~\cite{Ayon-Beato:1998hmi,Ayon-Beato:1999qin,Ayon-Beato:1999kuh,Ayon-Beato:2000mjt,Bronnikov:2000vy,Dymnikova:2004zc,Bronnikov:2022ofk,Balart:2014cga,Rodrigues:2018bdc} typically violating standard energy conditions~\cite{Zaslavskii:2010qz,Lan:2023cvz,Balart:2023odm,Borissova:2025hmj,Borissova:2025msp}, although it is seldom remarked that this regularization only works for a fine-tuned value of the black-hole mass. While less known, the effective descriptions of gravitational collapse processes into spherically symmetric regular black holes such as the ones considered in~\cite{Ziprick:2010vb,Taves:2014laa} based on lower-dimensional scalar-tensor theories, can regularize the Schwarzschild singularity for all values of the black-hole mass~\cite{Kunstatter:2015vxa}. The lower-dimensional theories considered in these effective descriptions are equivalent to subclasses of two-dimensional Horndeski theories, as discussed in recent treatments of static and dynamical regular solutions~\cite{Carballo-Rubio:2025ntd,Boyanov:2025pes}. In this setup the constant parameter corresponding to the black-hole mass in vacuum solutions arises as an integration constant when solving the equations of motion of the two-dimensional theory.

The possibility of deriving regular black holes from purely gravitational theories is fundamentally distinct in higher dimensions, where static spherically symmetric regular black holes can arise as vacuum solutions to polynomial quasi-topological gravities including an infinite tower of curvature terms~\cite{Bueno:2024dgm}. Polynomial quasi-topological gravities~\cite{Oliva:2010eb,Myers:2010ru,Dehghani:2011vu,Cisterna:2017umf,Bueno:2019ltp,Bueno:2019ycr, Bueno:2022res,Bueno:2024dgm,Bueno:2024zsx,Bueno:2025qjk,Bueno:2025tli,Ling:2025ncw,Frolov:2025ddw,Frolov:2026rcm} are polynomial higher-curvature extensions of the Einstein-Hilbert action which possess second-order field equations on spherically symmetric backgrounds and admit single-function static spherically solutions characterized entirely by an algebraic equation involving a one-variable function, which can be derived from the reduced Lagrangian on this type of backgrounds~\cite{Bueno:2025qjk}. Such theories do not exist in four spacetime dimensions~\cite{Bueno:2019ltp,Moreno:2023rfl}. However, there exist non-polynomial four-dimensional gravitational theories with similar properties,  such as the ones considered in~\cite{Deser:2007za,Gao:2012fd,Colleaux:2015yta,Colleaux:2017ibe,Colleaux:2019ckh,Bueno:2025zaj, Tan:2025hht}. In particular, \cite{Bueno:2025zaj} has constructed four-dimensional non-polynomial quasi-topological gravities whose static spherically symmetric solution space can reproduce the four-dimensional analogues of some of the regular-black hole spacetimes obtained from polynomial quasi-topological gravities in higher dimensions~\cite{Bueno:2024dgm}. However, this construction is limited in that the reduced actions on static spherically symmetric backgrounds, in all cases particular realizations of two-dimensional Horndeski actions, are a strictly smaller subset of the Horndeski actions that can be generated by polynomial quasi-topological gravities in higher dimensions. These Horndeski actions involve only polynomial functions of one of the primary invariants characterizing the Riemann on spherically symmetric backgrounds. By contrast, here we will consider genuinely non-polynomial gravities with second-order equations of motion on spherically symmetric backgrounds, i.e., actions which are built non-polynomially from curvature invariants, and for which the resulting reduced actions may involve non-polynomial functions of the aforementioned invariant. As a result, we do not encounter constraints on the free functions in the resulting subclass of Horndeski actions which arise in higher-dimensional polynomial quasi-topological gravities~\cite{Bueno:2025qjk} or in the four-dimensional non-polynomial quasi-topological gravities constructed in~\cite{Bueno:2025zaj}. This allows us to generate an extended class of regular black-hole spacetimes from four-dimensional gravities as opposed to~\cite{Bueno:2025zaj}, and moreover spacetimes that cannot be obtained from resummation series of curvature invariants as in~\cite{Bueno:2024dgm,Bueno:2025zaj}.\\
  
Concretely, we will construct non-polynomial four-dimensional gravities possessing second-order equations of motion on generic spherically symmetric backgrounds. To that end, we start from the most general subclass of two-dimensional Horndeski actions for a  two-dimensional Lorentzian metric and a scalar field which can be reached by the reduction of a four-dimensional Lagrangian density built from Riemann curvature invariants. The two independent degrees of freedom in the Horndeski action can be combined to form the line element of four-dimensional warped-product spacetimes~\cite{Carballo-Rubio:2025ntd}.

By lifting the elements  defining these two-dimensional Horndeski actions, to generally covariant four-dimensional densities, we can achieve that the resulting four-dimensional non-polynomial gravitational actions possess second-order field equations when evaluated on warped-product spacetimes, i.e., in particular when evaluated on generic spherically symmetric spacetimes. This in turn allows us to interpret some of the effective regular black-hole spacetimes which arise from two-dimensional Horndeski theories~\cite{Carballo-Rubio:2025ntd,Boyanov:2025pes} as actual solutions to a four-dimensional generally covariant gravitational theory. In particular, the relevant subspace of two-dimensional Horndeski actions is genuinely larger than the one spanned by the reduction of polynomial quasi-topological gravities~\cite{Bueno:2025qjk}. \\

The resulting non-polynomial gravities will in general admit static spherically symmetric solutions with two distinct functions. We will refer to {\it non-polynomial quasi-topological gravities} as the subset of these which admit single-function static spherically symmetric solutions. In particular, we will show that such solutions are characterized in general by an algebraic equation involving two functions of one of the primary Riemann invariants, which may be viewed as a generalization of the algebraic equation arising in polynomial quasi-topological gravities~\cite{Bueno:2025qjk} --- which involves only one of these functions.

Subsequently, as a specific example, we will focus on a restricted subset of non-polynomial quasi-topological gravities parametrized by a single one-variable function as in higher-dimensional polynomial quasi-topological gravities~\cite{Bueno:2025qjk}, and in restricted examples of the four-dimensional non-polynomial quasi-topological gravities considered in~\cite{Bueno:2025zaj}, which, however, are still polynomial in what concerns the form of their reduced actions. Hence, we view these theories as (subclasses of) genuine non-polynomial quasi-topological gravities. In particular, as a result, the four-dimensional analogues of the regular black-hole spacetimes constructed in higher-dimensional polynomial quasi-topological gravities in~\cite{Bueno:2024dgm}, and spacetimes that could not be constructed in restricted four-dimensional non-polynomial quasi-topological gravities in~\cite{Bueno:2025zaj}, can be obtained from purely gravitational actions in four dimensions.\\

In addition, we will establish connections between the four-dimensional non-polynomial gravities constructed and statements recently derived for higher-dimensional polynomial quasi-topological gravities~\cite{Bueno:2025qjk}. Some of the proofs provided therein make use of the reduced actions on spherically symmetric backgrounds, which in all cases are two-dimensional Horndeski actions of the type considered here, such that some of the results in~\cite{Bueno:2025qjk} hold regardless of which generally covariant gravitational actions the reduced actions are obtained from. On the other hand, we will also see that the assumption about polynomiality of the action for polynomial quasi-topological gravities implies that the majority of statements need to be extended in order to apply to non-polynomial gravities with second-order equations of motion on spherically symmetric backgrounds. In essence, the non-polynomial gravities constructed here can be subdivided into non-polynomial quasi-topological gravities admitting single-function static spherically symmetric solutions as well as another class of non-polynomial gravities, still with second-order equations of motion of spherically symmetric backgrounds, but for which static spherically symmetric solutions feature in general two distinct functions. \\

The remainder of this article is structured as follows. In Section~\ref{Sec:2DHorndeski4DGravity}, we consider which subclasses of two-dimensional Horndeski actions can be reached by a reduction of four-dimensional non-polynomial gravities and provide an explicit construction of non-polynomial gravitational actions which produce this subset of Horndeski actions. To that end, Subsection~\ref{SecSub:ActionEOM} reviews two-dimensional Horndeski theory, the equations of motion, and its interpretation as an effective theory for higher-dimensional warped-product spacetimes. Subsection~\ref{SecSub:Action4DReduced} specifies the  four-dimensional gravitational actions considered, as well as their reduction on such backgrounds. Subsection~\ref{SecSub:Invariants} contains a detailed discussion of polynomial curvature invariants for warped-product spacetimes. Therefrom, in Subsection~\ref{SecSub:Subclass}, we state the general form of Horndeski actions which can arise from the reduction of non-polynomial gravities with second-order equations of motion on warped-product backgrounds. In Subsection~\ref{SecSub:Expressing2DElementsAs4DCovariantDensities}, we express the building elements of subclasses of two-dimensional Horndeski actions as generally covariant four-dimensional densities, and thereby provide an explicit construction of four-dimensional gravitational actions which can produce any such given Horndeski action. Section~\ref{Sec:SSSSolutionSpace} analyzes single-function static spherically symmetric solutions of these theories and to that end establishes the necessary constraints on the reduced action allowing for such solutions. The algebraic equation determining the single metric function for these, so-called, non-polynomial quasi-topological gravities, involves two theory-dependent single-variable functions. For a subclass of these theories, this algebraic equation reduces to an equation analogous to the one in polynomial quasi-topological gravities. These contents are discussed in Subsections~\ref{SecSub:HorndeskiOneFunction} and~\ref{SecSub:EOMnf}. In Subsection~\ref{SecSub:ConnectionQTG}, we establish connections of these results to polynomial quasi-topological gravities.  In Section~\ref{Sec:RegularBHs}, we consider examples of regular black-hole spacetimes that lie in the solution space of non-polynomial quasi-topological gravities, and for which the corresponding gravitational actions can be stated explicitly. Concretely, in Subsections~\ref{SecSub:Hayward} and~\ref{SecSub:Dymnikova} we show how the Hayward and Dymnikova solutions can be obtained. Subsection~\ref{SecSub:OtherRBHs} collects further examples of regular black-hole spacetimes as solutions to non-polynomial quasi-topological gravities. We finish with a discussion in Section~\ref{Sec:Discussion}.

\section{From 2D Horndeski theories to 4D non-polynomial gravities}\label{Sec:2DHorndeski4DGravity}

\subsection{2D Horndeski action and equations of motion}\label{SecSub:ActionEOM}

In the following, we will review two-dimensional Horndeski theory and corresponding equations of motion as a second-order effective theory for the degrees of freedom of higher-dimensional warped-product spacetimes~\cite{Carballo-Rubio:2025ntd}. The Horndeski action~\cite{Horndeski:1974wa} for a two-dimensional Lorentzian metric $q_{ab}(y)$ and scalar field $\varphi(y)$ parametrized in terms of coordinates $\qty{y^a}_{a=1,2}$ can be written as~\cite{Kobayashi:2011nu}
\ba\label{eq:SHorndeski}
	S_{\text{Horndeski}}[q,\varphi] &=& \int \dd[2]{y} \sqrt{-q}\, \,\bigg[h_2(\varphi,\chi)- h_3(\varphi,\chi)\Box\varphi  + h_4(\varphi,\chi) {\mathcal{R}} \nn\\
	&{}& \quad \quad - 2 \partial_\chi h_4(\varphi,\chi)\qty[\qty(\Box \varphi)^2 - \nabla_a \nabla_b \varphi \nabla^a \nabla^b \varphi]\bigg]\,.
	\ea
	Here $\nabla^a$ denotes the covariant derivative of the metric $q_{ab}$ and $\mathcal{R}$ the Ricci scalar. Moreover, $\chi = \nabla_a \varphi \nabla^a \varphi$ is the kinetic term of the scalar field  and the functions $h_i$ are general functions of $\varphi$ and $\chi$. Horndeski theory is the most general theory with equations of motion of up to second order in the derivatives of $q_{ab}$ and $\varphi$, see, e.g., \cite{Kobayashi:2019hrl} for a review.~\footnote{Expressed in the form~\eqref{eq:SHorndeski}, the action is overparametrized. It is possible to recast the last two terms, given by the quartic two-dimensional Galileon Lagrangian~\cite{Deffayet:2011gz,Kobayashi:2011nu}, into the action for kinetic gravity braiding~\cite{Deffayet:2010qz} by means of partial integrations, see, e.g., \cite{Takahashi:2018yzc,Colleaux:2019ckh}. 
		Thus, two-dimensional Horndeski theory is 
		parametrized by two only free functions. This is particularly obvious at the level of the equations of motion~\eqref{eq:EOM1} and~\eqref{eq:EOM2}. Here, we take the action expressed in the form~\eqref{eq:SHorndeski} as a starting point, as this form is particularly convenient for establishing relations to reduced higher-dimensional gravities. \label{Footnote:2DHorndeski}
	}
The corresponding variations can be expressed as~\cite{Carballo-Rubio:2025ntd}
\ba
\mathcal{E}_{ab} &=& \frac{1}{\sqrt{-q}} \frac{\var{S}}{\var{q}^{ab}} \,=\, \beta \nabla_a \nabla_b \varphi - \qty(\frac{1}{2}\alpha +\beta \Box \varphi)q_{ab} + F \nabla_a \varphi \nabla_b \varphi\,, \label{eq:EOM1}\\
\mathcal{E}_\varphi &=&  \frac{1}{\sqrt{-q}} \frac{\var{S}}{\var\varphi} \,=\, - \beta \mathcal{R} + \partial_\varphi \alpha - 2 \chi \partial_\varphi F - 2 F \Box \varphi + 2 \partial_\varphi \beta \Box \varphi \nn\\
&{}&  \quad \quad \quad \quad \quad \,\,\,\,+2 \partial_\chi \beta \qty[\qty(\Box \varphi)^2 - \nabla_a \nabla_b \varphi \nabla^a \nabla^b \varphi ] - 2 \partial_\chi F \nabla_a \varphi \nabla^a \chi\,,\label{eq:EOM2}
\ea
where 
\ba
\alpha(\varphi,\chi) &=& h_2 + \chi \partial_\varphi \qty(h_3 - 2 \partial_\varphi h_4)\,,\label{eq:Alpha}\\
\beta(\varphi,\chi) &=& \chi \partial_\chi \qty(h_3 - 2 \partial_\varphi h_4) - \partial_\varphi h_4\,,\label{eq:Beta}\\
F(\varphi,\chi) &=& \partial_\chi \alpha - \partial_\varphi \beta \,.
\ea
The offshell Bianchi identity,
\ba\label{eq:Bianchi}
\nabla^a \mathcal{E}_{ab} + \frac{1}{2} \mathcal{E}_\varphi \nabla_b \varphi &=& 0\,,
\ea
resulting from general covariance of the action~\eqref{eq:SHorndeski}, implies that the equation of motion for the scalar field,   $\mathcal{E}_\varphi = 0$, is automatically satisfied onshell on the equations of motion for the metric, $\mathcal{E}_{ab} = 0$.\\

The field equations of two-dimensional Horndeski theory provide an effective framework to study higher-dimensional warped-product spacetimes in vacuum. To that end, the metric $q_{ab}(y)$ and scalar field $\varphi(y)$ are combined into the $d$-dimensional line element
\ba\label{eq:Metric}
g_{\mu\nu}(x) \dd{x^\mu} \dd{x^\nu} &=& q_{ab}(y) \dd{y}^a \dd{y}^b + \varphi(y)^2 \dd{\Omega}_{d-2}^2 \,.
\ea
Here $\dd{\Omega}^2_{d-2} = \gamma_{ij}(\theta) \dd{\theta}^i \dd{\theta}^j$ is the surface element of the $(d-2)$-dimensional unit sphere, with Euclidean metric $\gamma_{ij}(\theta)$ parametrized in terms of angular coordinates $\qty{\theta^i}_{i=1}^{d-2} $. I.e., we consider here the spherical reduction, but more generally $\dd{\Omega}^2_{d-2}$ may be replaced by the surface element of a $d-2$ dimensional compact space of constant sectional curvature. In four spacetime dimensions, we will write  the area element of the unit two-sphere as $\dd{\Omega}_2^2  = \dd{\theta}^2 + \sin^2(\theta) \dd{\phi}^2$.  

The variations~\eqref{eq:EOM1}--\eqref{eq:EOM2} can be used to construct a generalization of the $d$-dimensional Einstein tensor $G_{\mu\nu}$, applicable to warped-product spacetimes~\eqref{eq:Metric} and defined as~\cite{Carballo-Rubio:2025ntd}
\ba\label{eq:GGeneralized}
\mathbb{G}_{\mu\nu}(q,\varphi) &=& \varphi^{2-d}\mathcal{E}_{ab}\delta^a_\mu \delta^b_\nu - \frac{1}{2 (d-2)}\varphi^{5-d} \mathcal{E}_\varphi \gamma_{ij}\delta^i_\mu \delta^j_\nu\,.
\ea
This tensor is symmetric, of second-order in derivatives and covariantly conserved, as a result of the generalized Bianchi identity~\eqref{eq:Bianchi}: $\nabla^\mu \mathbb{G}_{\mu\nu} = 0$, where $\nabla^\mu$ denotes the covariant derivative of the $d$-dimensional spacetime metric~\eqref{eq:Metric}. In other words, within the subspace of warped-product spacetimes~\eqref{eq:Metric}, the  vacuum field equations of general relativity can be extended to  effective vacuum field equations~\cite{Carballo-Rubio:2025ntd}
\ba\label{eq:FieldEq}
\mathbb{G}_{\mu\nu} &=& 0\,.
\ea
These  equations can be used to describe geometries beyond the Schwarzschild solution to the vacuum Einstein equations, $G_{\mu\nu} =0$, restricted to warped-product spacetimes~\eqref{eq:Metric}. Examples are geometries describing regular black holes and horizonless compact objects, see, e.g.,~\cite{Borissova:2026dlz} for an application to renormalization-group improved black-hole spacetimes. For instance, the $d$-dimensional Hayward metric~\cite{Hayward:2005gi},
\ba
g_{\mu\nu}(x)\dd{x}^\mu \dd{x}^\mu &=& - f(r)\dd{t}^2 + \frac{\dd{r}^2}{f(r)} + r^2 \dd{\Omega}_{d-2}^2\,,\,\,\,\quad  \quad \,\,\, f(r)=1-\frac{2 M r^2}{r^{d-1} + 2 M l^{d-2}}\,,
\ea
with Arnowitt-Deser-Misner mass $M$ and regularization length parameter $l$, is a solution to the effective vacuum field equations~\eqref{eq:FieldEq} with functions $\alpha$ and $\beta$ given by~\cite{Boyanov:2025pes}
\ba
\alpha(\varphi,\chi) &=& \qty(d-2) l^2 (1-\chi)\varphi^{d-2}\qty[\qty(d-3)l^{2}\varphi^2 - \qty(d-1)l^d (1-\chi)] \qty[l^2 \varphi^2 - l^d (1-\chi)]^{-2}\,,\label{eq:AlphaHaywardd}\\
\beta(\varphi,\chi) &=& - \qty(d-2)l^4 \varphi^{d+1}  \qty[l^2 \varphi^2 - l^d (1-\chi)]^{-2}\,,\label{eq:BetaHaywardd}
\ea
see also~\cite{Borissova:2026dlz,Kunstatter:2015vxa}. Yet, so far it has remained unclear whether  effective  spacetimes obtained in this way can generically arise as actual solutions to a generally covariant $d$-dimensional gravitational theory. 

In the following, we will address this question exclusively for four-dimensional spacetimes. Concretely, we will demonstrate that some  classes of effective static spherically symmetric black-hole spacetimes in the solution space of the Horndeski equations of motion can indeed arise as vacuum solutions to four-dimensional  actions for a spacetime metric $g_{\mu\nu}(x)$. To that end, we will provide an exemplary explicit construction of  non-polynomial  four-dimensional gravities starting from corresponding restricted subclasses of two-dimensional Horndeski theories, and analyze the corresponding static spherically symmetric solution space.

\subsection{Reduced 4D action on warped-product backgrounds}\label{SecSub:Action4DReduced}

We will consider  four-dimensional gravitational actions built from contractions of the Riemann tensor and the spacetime metric in the form
\ba\label{eq:S}
S[g] &=& \int \dd[4]{x} \sqrt{-g}\,\mathcal{L}\qty(g^{\mu\nu},R_{\alpha\beta\gamma\delta})\,.
\ea
The Lagrangian density is assumed to be a function of scalar polynomial curvature invariants without covariant derivatives.
We do not assume $\mathcal{L}$ to be an analytic function of such invariants.
The reduced  action obtained by evaluating~\eqref{eq:S} on warped-product backgrounds~\eqref{eq:Metric}, is given by
\ba
\label{eq:SRed}
S[q,\varphi]  &=&\int \dd[2]{y} \sqrt{-q}\, \varphi^2 \,\mathcal{L}(q,\varphi)  \,\,\, \quad \text{where} \quad \,\,\, 
\mathcal{L}(q,\varphi)  \,\,=  \,\,\eval{\mathcal{L}}_{\eqref{eq:Metric}} \,.
\ea
Here we have absorbed a factor given by the area $\Omega_2$
of the unit two-sphere into the normalization on the left-hand side.
The equations of motion derived covariantly from~\eqref{eq:S} by variation with respect to $g^{\mu\nu}$ and restricted to warped-product spacetimes~\eqref{eq:Metric}, are equivalent to the Euler-Lagrange equations obtained by first inserting the ansatz~\eqref{eq:Metric} into~\eqref{eq:S} and varying the reduced action~\eqref{eq:SRed} with respect to  $q^{ab}$ and $\varphi$. This is the statement of the principle of symmetric criticality~\cite{Palais:1979rca,Fels:2001rv,Deser:2003up}, which in this case applies due to the truncation to the invariant sector under the isometry group of spatial rotations.

There exist gravitational theories~\eqref{eq:S} for which the reduced action~\eqref{eq:SRed} can be identified as a particular subclass of two-dimensional Horndeski actions for the  metric $q_{ab}(y)$ and scalar field $\varphi(y)$, written in the form~\eqref{eq:SHorndeski}. 
An example is the reduced action~\eqref{eq:SRed} for general relativity, 
\ba\label{eq:SGR}
S_{\text{GR}}[q,\varphi] &=& \int \dd[2]{y} \sqrt{-q}\, \varphi^2  \eval{R}_{\eqref{eq:Metric}} \,\,=\,\, \int \dd[2]{y} \sqrt{-q}\,  \qty[ \varphi^2 \mathcal{R} - 4 \varphi \Box \varphi +  2\qty( 1-\chi)] \,,
\ea
which amounts to the choice  $h_2(\varphi,\chi) = 2 (1-\chi)$, $h_3(\varphi,\chi) = 4 \varphi $ and $h_4(\varphi,\chi) = \varphi^2$ in~\eqref{eq:SHorndeski}. Another example is the reduced action~\eqref{eq:SRed} consisting of the Gauss-Bonnet invariant
\ba
\mathcal{G} &=& R^2 - 4 R_{\mu\nu}R^{\mu\nu} + R_{\mu\nu\rho\sigma}R^{\mu\nu\rho\sigma}\,.
\ea
In this case,
\ba\label{eq:SGB}
S_{\text{GB}}[q,\varphi] &=& \int \dd[2]{y} \sqrt{-q} \,\varphi^2 \eval{\mathcal{G}}_{\eqref{eq:Metric}} \,\,= \,\, \int \dd[2]{y} \sqrt{-q} \, \qty[ 4 \qty(1-\chi)\mathcal{R} + 8 \qty[\qty(\Box \varphi)^2 - \nabla_a \nabla_b \varphi \nabla^a \nabla^b \varphi] ]\,,\quad \quad
\ea
such that the free functions in~\eqref{eq:SHorndeski} are identified as  $h_2(\varphi,\chi) = h_3(\varphi,\chi) = 0$ and $h_4(\varphi,\chi) = 4(1-\chi)$. It is well-known that the theory constructed as a linear combination of the Ricci scalar and the Gauss-Bonnet invariant in four spacetime dimensions is in fact just Lovelock gravity~\cite{Lanczos:1938sf,Lovelock:1970zsf,Lovelock:1971yv}.~\footnote{The spacetime integral over the four-dimensional Gauss-Bonnet invariant is usually referred to as a topological term, which does not contribute to the equations of motion. Here, this can be seen by noticing that the functions $\alpha$ and $\beta$ computed from $h_2=h_3 = 0$ and $h_4 = 4(1-\chi)$ according to~\eqref{eq:Alpha}--\eqref{eq:Beta} are identically zero. However, there exist proposals for including effects of this term on the field equations in four dimensions by considering the field equations for Einstein-Gauss-Bonnet gravity in dimensions $d > 4$ and performing a $d$-dependent rescaling of the Gauss-Bonnet coupling, such that taking the limit $d \to 4$ results in a non-vanishing contribution from Gauss-Bonnet invariant to the trace equation~\cite{Glavan:2019inb}. See, e.g., \cite{Fernandes:2020rpa,Konoplya:2020qqh,Ghosh:2020vpc,Ghosh:2020syx,Konoplya:2020juj,Kumar:2020owy,Kumar:2020uyz,Kumar:2020bqf,Konoplya:2020cbv,Malafarina:2020pvl,Yang:2020jno,Feng:2020duo} for follow-up works, and~\cite{Gurses:2020ofy,Gurses:2020rxb,Arrechea:2020evj,Arrechea:2020gjw,Bonifacio:2020vbk,Ai:2020peo,Mahapatra:2020rds,Hohmann:2020cor,Cao:2021nng} for criticisms of this approach. A review on four-dimensional Einstein-Gauss-Bonnet gravity can be found in~\cite{Fernandes:2022zrq}.} The latter is the most general theory possessing second-order field equations on generic backgrounds.  However, on the reduced space of warped-product backgrounds~\eqref{eq:Metric}, two-dimensional Horndeski theory suggests that there are extended classes of four-dimensional gravitational theories yielding second-order equations of motion on these backgrounds, provided the corresponding Horndeski actions for the two-dimensional metric and scalar field can arise from the reduction of four-dimensional actions for the spacetime metric, such as~\eqref{eq:S}. In the following, we will show that by allowing the action to depend non-polynomiality on curvature invariants, indeed extended subclasses of two-dimensional Horndeski actions~\eqref{eq:SHorndeski} can be reached. Subsequently, we will  show that further restricted subclasses of these theories admit single-function static spherically symmetric solutions and thereby may be referred to as non-polynomial quasi-topological gravities, in analogy to polynomial quasi-topological gravities in higher dimensions~\cite{Oliva:2010eb,Myers:2010ru,Dehghani:2011vu,Cisterna:2017umf,Bueno:2019ltp,Bueno:2019ycr, Bueno:2022res,Bueno:2024dgm,Bueno:2024zsx,Bueno:2025qjk,Bueno:2025tli}. We will also explicitly establish in what sense the solution space of non-polynomial quasi-topological gravities is extended compared to the one of polynomial quasi-topological gravities.

\subsection{Curvature invariants for warped-product spacetimes}\label{SecSub:Invariants}

We start by noticing that the components of the Riemann tensor evaluated for warped-product metrics~\eqref{eq:Metric} can be written as
\ba\label{eq:riem_comps}
\eval{R\indices{_a_b_c_d}}_{\eqref{eq:Metric}}\,\, =\,\, \mathcal{R}\indices{_a_b_c_d}\,, \quad\,\,\,
\eval{R\indices{_a_i_b_j}}_{\eqref{eq:Metric}} \,\,=\,\, - \varphi \nabla_a\nabla_b \varphi \gamma_{ij} \,,\quad \,\,\,
\eval{R\indices{_i_j_k_l}}_{\eqref{eq:Metric}} \,\,=\,\, r^2\qty(1- \chi )\qty(\gamma_{ik} \gamma_{jl} - \gamma_{il} \gamma_{jk})\,,\,\,\,\,\,\,
\ea
where $\chi = \nabla_\mu \varphi \nabla^\mu \varphi =  \nabla_a \varphi \nabla^a \varphi $. We denote by $\mathcal{R}_{abcd}$, $\mathcal{R}_{ab}$ and $\mathcal{R}$ denote the Riemann tensor, Ricci tensor and Ricci scalar of the two-dimensional metric $q_{ab}$. Notice that here we are dealing specifically with $2+2$ warped-product spacetimes. In particular the Riemann tensor $\mathcal{R}_{abcd}$ is composed solely out of the trace component.
The components of the Ricci tensor are given by
\ba\label{eq:RicciTensor}
\eval{R_{ab}}_{\eqref{eq:Metric}} \,\,= \,\,\mathcal{R}_{ab} - 2 \frac{\nabla_a\nabla_b \varphi}{\varphi}\,,\quad \,\,\,
\eval{R_{ij}}_{\eqref{eq:Metric}} \,\,=\,\, \qty[-\varphi \Box \varphi + \qty(1-\chi )]\gamma_{ij}\,.
\ea
The Ricci tensor is block-diagonal, as the metric. The Ricci scalar reads
\ba\label{eq:R}
\eval{R}_{\eqref{eq:Metric}} &=& \mathcal{R} - 4  \frac{\Box \varphi}{\varphi} +  2 \frac{1-\chi}{\varphi^2}\,.
\ea
Therefrom we obtain the components of the Weyl tensor 
\ba\label{eq:WeylTensor}
\eval{C_{abcd}}_{\eqref{eq:Metric}} \,\,= \,\,\frac{1}{3}\Omega q_{a[c} q_{d]b}\,,\quad \,\,\,
\eval{C_{aibj}}_{\eqref{eq:Metric}} \,\,=\,\, - \frac{1}{12} \Omega  \varphi^2 q_{ab} \gamma_{ij}\,,\quad \,\,\,
\eval{C_{ijkl}}_{\eqref{eq:Metric}} \,\,= \,\,\frac{1}{3} \Omega \varphi^4 \gamma_{i[k}\gamma_{l]j}\,,
\ea
where 
\ba\label{eq:Omega}
\Omega &=& \mathcal{R} + 2 \frac{\Box \varphi}{\varphi} + 2 \frac{1-\chi}{\varphi^2}\,.
\ea
Due to the block-diagonal and explicit form of the inverse metric,
\ba
\qty(g^{\mu\nu}) &=& \text{diag}\qty{q^{ab}\,, \,\frac{1}{\varphi^2}\gamma^{ij} }\,,
\ea
all Riemann curvature invariants are functions of the four combinations
\ba\label{eq:BuildingBlocks}
\qty{\mathcal{R}\,,\, \frac{\Box \varphi}{\varphi}\,,\, \frac{ \nabla_a \nabla_b \varphi}{\varphi} \frac{ \nabla^a \nabla^b \varphi}{\varphi} \,,\,\frac{1-\chi}{\varphi^2}} &\equiv & \qty{\mathcal{R}\,,\, \eta\,,\, \tau\,,\,\psi} \,. 
\ea
Even though this can be seen explicitly by considering all possible scalar combinations that can be built from the components of the Riemann tensor in Eq.~\eqref{eq:riem_comps}, we can alternatively show it explicitly by considering the polynomial basis for Riemann invariants of a general four-dimensional spacetime consisting of the Zakhary-McIntosh (ZM) invariants~\cite{Carminati:1991ddy,Zakhary:1997xas},
\begin{center}
	\begin{minipage}{0.4\textwidth}
		\ba
		\mathcal{I}_{R}  &=& R\label{eq:ZMFirst}\,,\\
		\mathcal{I}_{R^2}  &=& R\indices{_\mu^\nu}R\indices{_\nu^\mu} \,,\\
		\mathcal{I}_{C^2} &=& C_{\mu\nu\rho\sigma}C^{\mu\nu\rho\sigma}\,,\\
		\mathcal{I}_{C\tilde{C}}  &=& C_{\mu\nu\rho\sigma}\tilde{C}^{\mu\nu\rho\sigma}\,,\\
		\mathcal{I}_{R^3}  &=& R\indices{_\mu^\nu}R\indices{_\nu^\rho} R\indices{_\rho^\mu}\,,\\
		\mathcal{I}_{R^2 C}  &=& R\indices{^\mu^\nu}R\indices{^\rho^\sigma} C_{\rho\mu\nu\sigma} \label{eq:ZMR2C}\,,\\
		\mathcal{I}_{R^2 \tilde{C}}  &=& R\indices{^\mu^\nu}R\indices{^\rho^\sigma} \tilde{C}_{\rho\mu\nu\sigma}\,,\\
		\mathcal{I}_{C^3}  &=& C\indices{_\mu_\nu ^\rho ^\sigma} C\indices{_\rho_\sigma^\alpha^\beta}C\indices{_\alpha_\beta^\mu^\nu} \,,\\
		\mathcal{I}_{C^2 \tilde{C}}  &=&  C\indices{_\mu_\nu ^\rho ^\sigma} C\indices{_\rho_\sigma^\alpha^\beta}\tilde{C}\indices{_\alpha_\beta^\mu^\nu} \,,\\
		&{}&\nn
		\ea
	\end{minipage}
	\hspace{0.5cm}
	\begin{minipage}{0.5\textwidth}
		\ba
		\mathcal{I}_{R^4}  &=& R\indices{_\mu^\nu}R\indices{_\nu^\rho} R\indices{_\rho^\sigma} R\indices{_\sigma^\mu}\,,\\
		\mathcal{I}_{R^2 C^2}  &=& R\indices{^\mu^\nu} R\indices{_\rho_\sigma}C\indices{_\alpha _\mu_\nu _\beta} C\indices{^\alpha^\rho^\sigma^\beta} \,,\label{eq:R2C2}\\
		\mathcal{I}_{R^2 \tilde{C}^2}  &=& R\indices{^\mu^\nu} R\indices{_\rho_\sigma} \tilde{C}\indices{_\alpha _\mu_\nu _\beta} \tilde{C}\indices{^\alpha^\rho^\sigma^\beta} \,,\\
		\mathcal{I}_{R^2 C\tilde{C}}  &=& R\indices{^\mu^\nu} R\indices{_\rho_\sigma}C\indices{_\alpha _\mu_\nu _\beta}\tilde{C}\indices{^\alpha ^\rho^\sigma ^\beta}\,,\\
		\mathcal{I}_{R^4 C}   &=& R\indices{_\mu ^\alpha} R\indices{_\alpha^\rho}  R\indices{_\nu ^\beta} R\indices{_\beta ^\sigma} C\indices{^\mu^\nu_\rho_\sigma}\,,\\
		\mathcal{I}_{R^4\tilde{C}}   &=& R\indices{_\mu ^\alpha} R\indices{_\alpha^\rho}  R\indices{_\nu ^\beta} R\indices{_\beta ^\sigma} \tilde{C}\indices{^\mu^\nu_\rho_\sigma}\,,\\
		\mathcal{I}_{R^2 C^3}    &=& R\indices{^\mu^\nu}R\indices{^\rho^\sigma}C\indices{^\alpha ^\beta^\gamma^\delta}C\indices{_\alpha _\mu_\nu _\delta}C\indices{_\beta_\rho_\sigma_\gamma} \,, \label{eq:R2C3}\\
		\mathcal{I}_{R^2 \tilde{C}^3}  &=& R\indices{^\mu^\nu}R\indices{^\rho^\sigma}\tilde{C}\indices{^\alpha ^\beta^\gamma^\delta} \tilde{C}\indices{_\alpha _\mu_\nu _\delta} \tilde{C}\indices{_\beta_\rho_\sigma_\gamma} \label{eq:ZMLast}\,,\\
		&{}&\nn\\
		&{}&\nn
		\ea
	\end{minipage}
	\hspace{0.1cm}
\end{center}
see also, e.g., \cite{Overduin:2020aiq,Held:2021vwd,Borissova:2023kzq}.
Here $\tilde{C}$ denotes the dual  Weyl tensor defined as $\tilde{C}_{\mu\nu\rho\sigma} = \frac{1}{2}\epsilon_{\mu\nu\alpha\beta}C\indices{^\alpha^\beta_\rho_\sigma}$. 
Any scalar polynomial Riemann invariant of order $n$ in the curvature can be written as a linear combination of products of ZM invariants such that the total power of curvatures in each summand equals $n$.~\footnote{A concrete prescription, based on symmetries and geometric identities satisfied by the Riemann tensor, allowing for the decomposition of a given polynomial Riemann invariant into a polynomial sum of ZM invariants, is a conceptual question that is out of the scope of this work.}
Note that the ZM invariants involve only contractions consisting of up to five curvature tensors. Moreover, in all invariants in which the indices of a Ricci tensor are contracted with the indices of a single Weyl tensor, we may replace the former by the traceless Ricci tensor
\ba
\hat{R}_{\mu\nu} &=& R_{\mu\nu} - \frac{1}{4} Rg_{\mu\nu} \,.
\ea
The ZM invariants evaluated for the warped-product metrics~\eqref{eq:Metric} are given in Appendix~\ref{App:ZMInvariants}. They are polynomial functions of the four quantities defined in~\eqref{eq:BuildingBlocks}. This shows explicitly that any Riemann curvature invariant of the metric~\eqref{eq:Metric} can be combined out of these elements.

\subsection{Restricted subclass of 2D Horndeski theories}\label{SecSub:Subclass}

As explained previously, any curvature invariant constructed polynomially from the Riemann tensor  for the metrics~\eqref{eq:Metric} is a function of the four quantities~\eqref{eq:BuildingBlocks}. When evaluating any four-dimensional generally covariant Lagrangian density~\eqref{eq:S} on this class of  backgrounds, the parts which contain only $\varphi$ and $\chi$ can depend on these two variables only  through the combination
\ba\label{eq:Psi}
\psi &=& \frac{1-\chi}{\varphi^2}\,.
\ea
Therefrom we conclude that the only two-dimensional Horndeski actions~\eqref{eq:SHorndeski} which can be generated from such a reduction, prior to performing any partial integrations, are those for which
\ba\label{eq:Hi}
H_2(\varphi,\chi) \,\,= \,\,\frac{h_2(\varphi,\chi)}{\varphi^2}\,,\quad H_3(\varphi,\chi) \,\,= \,\,\frac{h_3(\varphi,\chi)}{\varphi}\,,\quad H_4(\varphi,\chi) \,\,=\,\, \frac{h_4(\varphi,\chi)}{\varphi^2}
\ea
are functions of $\psi$ only, i.e.,
\ba\label{eq:SHorndeskiSubclass}
S_{\text{Horndeski}}[q,\varphi] &=& \int \dd[2]{y} \sqrt{-q}\, \varphi^2 \,\qty[H_2(\psi) - H_3(\psi)\eta + H_4(\psi) {\mathcal{R}} + 2 H_4'(\psi)\qty(\eta^2 - \tau)]\,,
\ea
where a prime denotes the derivative with respect to $\psi$.
With the above redefinitions of the free functions in the Horndeski action, the reduced action for general relativity~\eqref{eq:SGR} corresponds to the choice $H_2(\psi) = 2\psi$, $H_3(\psi) = 4$ and $H_4(\psi) = 1$.\\

Notice that if each of the elements~\eqref{eq:BuildingBlocks} could be lifted independently to a four-dimensional covariant density, depending in general non-polynomially on curvature invariants, then it would be possible to  construct four-dimensional gravities whose reduction can yield any Horndeski action in the subclass~\eqref{eq:SHorndeskiSubclass} and  the constraint $H_i = H_i(\psi)$ would be a sufficient condition for the existence of a four-dimensional generally covariant theory~\eqref{eq:S} which yields this Horndeski theory upon reduction.~\footnote{This constraint is however not necessary, as explained, since performing partial integrations in~\eqref{eq:SHorndeskiSubclass} in general does not preserve the functional dependence $H_i = H_i(\psi)$.} We will see that such a covariant expression of the elements~\eqref{eq:BuildingBlocks} is indeed possible by using non-polynomial combinations of curvature invariants. By constrast, this is not possible to achieve from polynomial gravities with second-order equations of motion on~\eqref{eq:Metric}, in which case the assumption about polynomiality of the action imposes a constraint relation among the functions $H_i(\psi)$, which in turn can be parametrized entirely in terms of a single function $H(\psi)$ as shown in~\cite{Bueno:2025qjk}. We will elaborate on this later.\\

In the following, it will be convenient to change variables
\ba\label{eq:VariableChange}
(\varphi,\,\chi)\quad \to \quad  (\varphi,\,\psi)\,, \quad \quad 
(\partial_\varphi,\,\partial_\chi)  \quad \to\quad \qty(\partial_\varphi - \frac{2 \psi}{\varphi}\partial_\psi,\, -\frac{1}{\varphi^2}\partial_\psi)\,.
\ea
As a result, and for future reference, we note that for the subclass of Horndeski theories with action~\eqref{eq:SHorndeskiSubclass}, the functions $\alpha$ and $\beta$ defined in~\eqref{eq:Alpha} and~\eqref{eq:Beta} are given by
\ba
\alpha(\varphi,\psi) &=&  \varphi^2 H_2- \qty(1-\varphi^2 \psi) \qty[- H_3 + 2 \psi H'_3+ 4 H_4 - 4 \psi H_4' + 8 \psi^2 H_4'']\,,\label{eq:AlphaVarphiPsi}\quad \\
\beta(\varphi,\psi) 
&=& - \frac{1}{\varphi}\qty(1- \varphi^2 \psi) \qty[H_3 - 4 H_4 + 4 \psi H_4']' - 2 \varphi H_4 + 2 \varphi \psi H_4'\,,\label{eq:Beta:VarphiPsi}
\ea
where $H_i = H_i(\psi)$. For example, in general relativity the terms in square brackets vanish, and in this case $\alpha(\varphi,\psi) =  2 \varphi^2 \psi$ and $\beta(\varphi,\psi) = - 2 \varphi$. We will consider more general spacetimes with this property later in Section~\ref{Sec:SSSSolutionSpace}. 
\\

Generically, there will be many inequivalent ways of expressing the four scalars in~\eqref{eq:BuildingBlocks} as covariant densities evaluated on~\eqref{eq:Metric}. One may, for instance, select any four independent equations out of the evaluated ZM invariants given in Appendix~\ref{App:ZMInvariants}, and solve for the elements $\mathcal{R}$, $\eta$, $\tau$ and $\psi$.
If explicit solutions exist, these will be in most cases non-polynomial functions of curvature invariants and linear combinations thereof raised to non-integer powers --- although by an appropriate choice of invariants used for the elimination, it possible to generate four-dimensional Lagrangian densities which are non-polynomial in curvature invariants, but do not contain non-integer powers of invariants. We shall provide an exemplary construction in the next Section~\ref{SecSub:Expressing2DElementsAs4DCovariantDensities}. Depending on the choice of invariants used when expressing the four elements~\eqref{eq:BuildingBlocks} covariantly, the outcomes of ``lifting" a given Horndeski action~\eqref{eq:SHorndeskiSubclass} to a gravitational action~\eqref{eq:S} will be different. In particular, the reduction of a four-dimensional gravitational action to yield a two-dimensional Horndeski theory, and the subsequent lifting back to a four-dimensional gravitational action, do not commute in general. However, despite the degeneracy in possible resulting four-dimensional Lagrangian densities, this approach provides a systematic procedure which demonstrates that any two-dimensional Horndeski actions~\eqref{eq:SHorndeskiSubclass} can arise from the reduction of four-dimensional gravitational actions of the form~\eqref{eq:S}. Our focus here is not the concrete realization of such a construction in terms of a complicated four-dimensional action. Instead, throughout we take the viewpoint that the possibility of obtaining Horndeski actions of the form~\eqref{eq:SHorndeskiSubclass} as a reduction of four-dimensional gravities provides a proof of principle that many regular black-hole spacetimes, effective solutions to two-dimensional Horndeski theories, can arise as vacuum solutions to pure gravity in four dimensions, as we shall discuss in detail in Sections~\ref{Sec:SSSSolutionSpace} and~\ref{Sec:RegularBHs}.

\subsection{Expressing 2D action elements as 4D covariant densities}\label{SecSub:Expressing2DElementsAs4DCovariantDensities}

In the following, we consider an exemplary construction of  four-dimensional non-polynomial gravitational actions~\eqref{eq:S} whose reduction yields any representative in the subclass of Horndeski actions~\eqref{eq:SHorndeskiSubclass}, subject to the requirement that no non-integer powers of curvature invariants appear in the resulting four-dimensional action. We emphasize that such type of construction is by no means unique, but rather serves the purpose of later explicitly stating gravitational actions that yield regular black-hole spacetimes as solutions.\\

To begin, we remind the reader according to~\eqref{eq:SGR}, or equivalently~\eqref{eq:RicciScalar}, that the Ricci scalar evaluated for the metrics~\eqref{eq:Metric} can be expressed as
\ba\label{eq:RicciScalarEvaluated}
\eval{\mathcal{I}_{R}}_{\eqref{eq:Metric}} &=& \mathcal{R} - 4 \eta + 2 \psi \,\,\, \equiv \,\, \rho\,,
\ea
where we have introduced the symbol $\rho$ to denote the evaluated Ricci scalar, in order to distinguish it from the Ricci scalar as a covariant density.
Next, we note that the evaluations of the squared and cubed Weyl invariants can be expressed in terms of the quantity
\ba\label{eq:OmegaConcrete}
\Omega &=& \mathcal{R} + 2\eta + 2 \psi\,,
\ea
defined in~\eqref{eq:Omega}, as
\ba\label{eq:IC2IC3}
\eval{\mathcal{I}_{C^2}}_{\eqref{eq:Metric}} \,\,= \,\,\frac{1}{3}\Omega^2\,,\,\,\,\quad 
\eval{\mathcal{I}_{C^3}}_{\eqref{eq:Metric}} \,\,= \,\,\frac{1}{18}\Omega^3\,.
\ea
More generally, $\eval{\mathcal{I}_{C^k}}_{\eqref{eq:Metric}} = 3^{-k} \qty[2 + (-2)^{2-k}]\Omega^k$ can be shown to hold for any integer $k \geq 2$, where $\mathcal{I}_{C^k}$ is the invariant constructed analogously from $k$ powers of the Weyl tensor~\cite{Deser:2005pc,Deser:2007za}. Thus, the quantity $\Omega$, which  in particular is linear in all four elements~\eqref{eq:BuildingBlocks}, can be expressed covariantly as a ratio of subsequent Weyl invariants evaluated on~\eqref{eq:Metric}. 
 Here, we proceed by expressing $\Omega$ in terms of covariant densities as follows,
\ba\label{eq:OmegaLift}
\Omega &=& \eval{6 \frac{\mathcal{I}_{C^3}}{\mathcal{I}_{C^2}}}_{\eqref{eq:Metric}}\,.
\ea
We will refer to such non-polynomial densities as being of overall first order in the curvature.
Combining the expression for $\Omega$ in~\eqref{eq:OmegaConcrete} with the expression for the evaluated Ricci scalar $\rho$ in~\eqref{eq:RicciScalarEvaluated}, we see that $\eta$ appearing in the Horndeski action~\eqref{eq:SHorndeskiSubclass}, can be eliminated  as
\ba
\eta &=&  -\frac{1}{6}\qty(\rho - \Omega)\,.
\ea
Thus, this element  can be lifted to the covariant density
\ba\label{eq:ILift}
\eta
&=&  \eval{\qty[-\frac{1}{6} \mathcal{I}_R  +\frac{\mathcal{I}_{C^3}}{\mathcal{I}_{C^2}}]}_{\eqref{eq:Metric}} \,\,
 \equiv \,\, \eval{\mathcal{H}}_{\eqref{eq:Metric}}\,.
\ea

In the next step, we demonstrate how the remaining elements $\mathcal{R}$, $\tau$ and $\psi$ appearing in the Horndeski action~\eqref{eq:SHorndeskiSubclass} can be lifted to four-dimensional covariant densities. To that end, it is convenient to introduce the two invariants built from the squared and cubed traceless Ricci tensor,
\ba
\mathcal{I}_{\hat{R}^2} &=& \hat{R}\indices{_\mu^\nu} \hat{R}\indices{_\nu^\mu}\,,\\
\mathcal{I}_{\hat{R}^3} &=& \hat{R}\indices{_\mu^\nu} \hat{R}\indices{_\nu^\rho} \hat{R}\indices{_\rho^\mu}\,.
\ea
Evaluated for the metrics~\eqref{eq:Metric}, these are given by
\ba
\eval{\mathcal{I}_{\hat{R}^2}}_{\eqref{eq:Metric}} &=& \frac{1}{4}\mathcal{R}^2 - \mathcal{R}\mathcal{\psi} - 2 \eta^2 + 4 \tau+ \mathcal{\psi}^2 \,\,\equiv \,\,\Sigma\,,\label{eq:IRTraceless2}\\
\eval{\mathcal{I}_{\hat{R}^3}}_{\eqref{eq:Metric}} &=& - \frac{3}{2}\mathcal{R} \eta^2 + 3 \mathcal{R}\tau +3 \eta^2\psi - 6 \tau\psi\,.\label{eq:IRTraceless3}
\ea
Note, in particular, that the latter is linear in all the elements~\eqref{eq:BuildingBlocks} which have not yet been expressed as covariant densities, i.e., in $\mathcal{R}$, $\tau$ and $\psi$. Thus, this invariant is particularly useful for eliminations without generating roots in the resulting expressions.\\

We proceed by observing that the ZM invariant~\eqref{eq:ZMR2C}, consisting of two Ricci tensors contracted with one Weyl tensor, can be expressed as 
\ba
\eval{\mathcal{I}_{R^2 C}}_{\eqref{eq:Metric}} &=&  -\Omega \qty[\qty(\eta^2 -2 \tau) + \frac{1}{3}\Sigma]  \,,
\ea
where $\Sigma$ was defined in~\eqref{eq:IRTraceless2} as the evaluated squared traceless Ricci invariant. Eliminating $\tau$ from the previous expression and making use of~\eqref{eq:OmegaLift} and~\eqref{eq:ILift}, we see that this quantity can be lifted to the covariant density~\footnote{Alternatively, $\tau$ can be eliminated from the ZM invariants~\eqref{eq:R2C2} or~\eqref{eq:R2C3}, consisting of two Ricci tensors contracted with two or three Weyl tensors, respectively. When evaluated on~\eqref{eq:Metric}, these can be expressed as
\ba
\eval{\mathcal{I}_{R^2 C^2}}_{\eqref{eq:Metric}} &=&  \Omega^2 \qty[\frac{1}{6}\qty(\eta^2 -2 \tau) + \frac{1}{9}\Sigma]  \,,\\
\eval{\mathcal{I}_{R^2 C^3}}_{\eqref{eq:Metric}} &=&  -\Omega^3 \qty[\frac{1}{12}\qty(\eta^2 -2 \tau) + \frac{1}{27}\Sigma]\,. 
\ea
Proceeding with either one of these eliminations will lead to generically inequivalent four-dimensional theories.
}
\ba\label{eq:JLift}
\tau &=&\eval{\qty[\frac{1}{72}\mathcal{I}_{R}^2 +\frac{1}{6} \mathcal{I}_{\hat{R}^2} - \frac{1}{6}\frac{\mathcal{I}_R \mathcal{I}_{C^3}}{\mathcal{I}_{C^2}} + \frac{1}{2}\frac{\mathcal{I}_{C^3}^2}{\mathcal{I}_{C^2}^2} + \frac{1}{12}\frac{\mathcal{I}_{C^2}\mathcal{I}_{R^2 C}}{\mathcal{I}_{C^3}}]}_{\eqref{eq:Metric}}  \,\, \equiv \,\, \eval{\mathcal{T}}_{\eqref{eq:Metric}}\,.
\ea

To derive an expression for $\psi$ as a four-dimensional density evaluated on~\eqref{eq:Metric}, we use that the evaluation of $\mathcal{I}_{\hat{R}^3}$ in~\eqref{eq:IRTraceless3} is linear in $\psi$. This holds still after inserting~\eqref{eq:RicciScalarEvaluated} for the two-dimensional Ricci scalar $\mathcal{R}$. Hence, solving for $\psi$, after using~\eqref{eq:ILift} and \eqref{eq:JLift}, we arrive at
\ba \label{eq:PsiLift}
\psi &=& \eval{\qty[\frac{1}{12}\mathcal{I}_R + \frac{\mathcal{I}_{C^3}}{\mathcal{I}_{C^2}} -  \frac{\mathcal{I}_{\hat{R}^3}\mathcal{I}_{C^3}}{\mathcal{I}_{C^2} \mathcal{I}_{R^2 C} + 2 \mathcal{I}_{\hat{R}^2} \mathcal{I}_{C^3}}]}_{\eqref{eq:Psi}}  \,\, \equiv \,\, \eval{\mathcal{P}}_{\eqref{eq:Metric}}\,.\\
\nn
\ea
Finally, after eliminating $\mathcal{R}$ from~\eqref{eq:RicciScalarEvaluated} and using~\eqref{eq:ILift} and~\eqref{eq:PsiLift}, we conclude that the two-dimensional Ricci scalar can be lifted to the four-dimensional density
\ba\label{eq:RLift}
\mathcal{R} &=& \eval{\qty[\frac{1}{ \mathcal{I}_{C^2} \mathcal{I}_{R^2C} + 2\mathcal{I}_{\hat{R}^2} \mathcal{I}_{C^3} }\qty[\frac{1}{6} \mathcal{I}_{R} \mathcal{I}_{C^2} \mathcal{I}_{R^2 C}  +  \frac{1}{3} \mathcal{I}_{R}\mathcal{I}_{\hat{R}^2}  \mathcal{I}_{C^3} + 4 \frac{ \mathcal{I}_{\hat{R}^2} \mathcal{I}_{C^3}^2}{\mathcal{I}_{C^2}} 	+ 2 \mathcal{I}_{\hat{R}^3} \mathcal{I}_{C^3} + 2  \mathcal{I}_{C^3} \mathcal{I}_{R^2 C}]]}_{\eqref{eq:Metric}} \nn\\
&\equiv &  \eval{\mathcal{K}}_{\eqref{eq:Metric}}\,.
\ea
In summary, the framework presented here allows for systematically constructing non-polynomial gravities with second-order equations of motion on warped-product spacetimes~\eqref{eq:Metric}, and serves as a proof of principle that any two-dimensional Horndeski action~\eqref{eq:SHorndeskiSubclass} can arise from the reduction of a four-dimensional gravity~\eqref{eq:S}. Specifically, following the exemplary construction above, the reduction of the following gravitational action,
\ba\label{eq:SLift}
S[g] &=& \int \dd[4]{x} \sqrt{-g}\, \qty[H_2(\mathcal{P}) - H_3(\mathcal{P}) \mathcal{H} + H_4(\mathcal{P})\mathcal{K} + 2  H_4'(\mathcal{P}) \qty(\mathcal{H}^2 - \mathcal{T}) ]\,,
\ea
produces any representantive in the subclass of Horndeski actions~\eqref{eq:SHorndeskiSubclass}. As a result, configurations $q_{ab}$ and $\varphi$ which lie in the solution space of these Horndeski theories can be associated with spacetimes $g_{\mu\nu}$ composed as in~\eqref{eq:Metric} which are, in fact, solutions to a four-dimensional generally covariant theory. Notice also that by taking powers of the covariant density $\mathcal{P}$ defined in~\eqref{eq:PsiLift}, one may generate four-dimensional higher-curvature actions of arbitrarily high order, which will have second-order equations of motion when evaluated on warped-product backgrounds~\eqref{eq:Metric}.\\

 We close this subsection by mentioning that the density appearing in the denominator of the last term in~\eqref{eq:PsiLift}, and in the denominator of the first factor in~\eqref{eq:RLift}, coincides with the density in the denominators of recently constructed four-dimensional non-polynomial quasi-topological gravities~\cite{Bueno:2025zaj}. The actions constructed in~\cite{Bueno:2025zaj} possess second-order equations of motion on warped-product backgrounds~\eqref{eq:Metric}. Their reduction yields Horndeski actions of the form~\eqref{eq:SHorndeskiSubclass} whereby the functions $H_i(\psi)$ are necessarily polynomial functions of $\psi$. By contrast, our construction here yields non-polynomial gravities with reduced actions~\eqref{eq:SHorndeskiSubclass} in which $H_i(\psi)$ can be arbitrary functions of $\psi$. In particular, these theories feature an extended solution space compared to the theories constructed in~\cite{Bueno:2025zaj}, as we shall see. 

\section{Single-function static spherically symmetric solutions}\label{Sec:SSSSolutionSpace}

In the following, we will analyze conditions for the subclass of Horndeski actions~\eqref{eq:SHorndeskiSubclass} which can be generated from the reduction of non-polynomial four-dimensional gravities, such as, e.g.,~from~\eqref{eq:SLift}, to admit single-function static spherically symmetric solutions~\footnote{This notion is meaningful for the four-dimensional spacetimes.} determined by an algebraic equation. We will refer to such classes of gravitational theories as {\it non-polynomial quasi-topological gravities}.

\subsection{Non-polynomial quasi-topological gravities}
\label{SecSub:HorndeskiOneFunction}

We will consider static configurations parametrized in terms of coordinates $\{y^a\} = \qty{t,r}$ and one free function $f$, in the form 
\ba\label{eq:AnsatzSSSn1}
q_{ab}(y)\dd{y}^a \dd{y}^b &=& -f(r)\dd{t}^2 + \frac{\dd{r}^2}{f(r)}\,\,\, \quad \text{and} \quad \,\,\, \varphi(y) \,\,=\,\, r\,.
\ea

As a result, the four-dimensional static spherically symmetric spacetimes composed according to~\eqref{eq:Metric}, i.e.,
\ba\label{eq:MetricSSSn1}
g_{\mu\nu}(x)\dd{x}^\mu \dd{x}^\nu &=& -f(r)\dd{t}^2 + \frac{\dd{r}^2}{f(r)} + r^2 \dd{\Omega}^2\,,
\ea
 will be solutions to a  generally covariant four-dimensional gravitational theory. \\

 For the ansatz specified in~\eqref{eq:AnsatzSSSn1}, it follows that $\chi = f$ onshell
and the two independent components of the Horndeski equations of motion~\eqref{eq:FieldEq} are given by
\ba
\mathbb{G}_{tt} &=&
 \frac{1}{r^2}\mathcal{E}_{tt}  \,\, = \,\, 
\frac{ f}{2 r^2}\qty[\alpha + \beta f'] \,\, =\,\,0 \,,\label{eq:EOMtt}\\
\mathbb{G}_{rr} &=& \frac{1}{r^2}\mathcal{E}_{rr} 
 \,\,= \,\,-\frac{1}{2  r^2  f}\qty[\alpha + \beta f'  - 2  f\qty(\partial_f \alpha - \partial_r \beta) ] \,\, =\,\, 0\,,\label{eq:EOMrr}
\ea
where still the representation $\alpha = \alpha(\varphi,\chi)$ and $\beta = \beta(\varphi,\chi)$ is used, with $\varphi =r$ and $\chi = f$ onshell.
The angular components of the generalized Einstein tensor $\mathbb{G}_{\mu\nu}$ do not give rise to further independent equations, as follows by using $\nabla^\mu \mathbb{G}_{\mu\nu} = 0$.
From~\eqref{eq:EOMtt} and~\eqref{eq:EOMrr}, we can state the independent equations of motion as follows,
\ba
\alpha + \beta f' &=& 0\,,\label{eq:Eq1}\\
\partial_f \alpha - \partial_r \beta &=& 0\label{eq:Eq2}\,.
\ea 
A solution to the second equation can be found in terms of a function $\Omega(\varphi,\chi)$ satisfying~\cite{Boyanov:2025pes,Carballo-Rubio:2025ntd,Borissova:2026dlz}
\ba
\alpha(\varphi,\chi) \,\, = \,\, \partial_\varphi \Omega(\varphi,\chi)\,\,\,\quad \,\,\,\text{and} \,\,\,\quad \,\,\, \beta(\varphi,\chi) \,\, = \,\, \partial_\chi \Omega(\varphi,\chi)\,.\label{eq:Omegarf}
\ea
In the following, we will perform the change of variables $(\varphi,\chi) \to (\varphi,\psi)$ detailed in~\eqref{eq:VariableChange} and require that the functions $\alpha(\varphi,\psi)$ and $\beta(\varphi,\psi)$ take the form~\eqref{eq:AlphaVarphiPsi} and~\eqref{eq:Beta:VarphiPsi}. This applies to the subclass of Horndeski actions~\eqref{eq:SHorndeskiSubclass}, i.e., when $H_i = H_i(\psi)$, which arise as reduced actions to the non-polynomial gravities constructed in Subsection~\ref{SecSub:Expressing2DElementsAs4DCovariantDensities}.  The equations of motion~\eqref{eq:Eq1}--\eqref{eq:Eq2} then become
\ba
\alpha + \beta f' &=& 0\,,\label{eq:Eq1Again}\\
-\frac{1}{r^2}\partial_\psi \alpha - \partial_r \beta + \frac{2 \psi}{r} \partial_\psi \beta &=& 0\,, \label{eq:Eq2Rewritten}
\ea
where now $\alpha=\alpha(\varphi,\psi)$ and $\beta=\beta(\varphi,\psi)$, with $\varphi = r$ and $\psi = \psi(r,f)$ onshell, and where
\ba
\psi(r,f) &=& \frac{1-f}{r^2}\,.
\ea
The defining relations~\eqref{eq:Omegarf} for the function $\Omega(\varphi,\psi)$  transform into
\ba
\alpha(\varphi,\psi)\,\,=\,\, \partial_\varphi \Omega(\varphi,\psi) - \frac{2 \psi}{\varphi} \partial_\psi \Omega(\varphi,\psi) \,\,\, \quad \,\,\, \text{and} \,\,\, \quad \,\,\,
\beta(\varphi,\psi) \,\,=\,\,
-\frac{1}{\varphi^2} \partial_\psi \Omega(\varphi,\psi) \,.\label{eq:Omegarpsi}
\ea
The functions $\alpha(\varphi,\psi)$ and $\beta(\varphi,\psi)$ given in~\eqref{eq:AlphaVarphiPsi}--\eqref{eq:Beta:VarphiPsi}  can be written as
	\ba
	\alpha(\varphi,\psi) &=& \varphi^2 F_1(\psi) + F_2(\psi)\,,\label{eq:alpha}\\
	\beta(\varphi,\psi) &=& \frac{1}{\varphi}F_3'(\psi) + \varphi F_4(\psi)\,,\label{eq:beta}
	\ea
	for functions $F_i(\psi)$ defined in terms of $H_i(\psi)$ and their derivatives as follows,
	\ba
	F_1 &=& H_2  - \psi F_2 \,,\label{eq:F1}\\
	F_2 &=& H_3 - 2 \psi  H'_3  - 4 H_4 + 4 \psi H_4' - 8 \psi^2 H_4''
	\,,\label{eq:F2}\\
	F_3 &=& -H_3 + 4 H_4 - 4 \psi H_4'\,, \label{eq:F3}\\
	F_4 &=& -\psi F_3' - 2 H_4 + 2 \psi H_4' \,. \label{eq:F4}
	\ea
	For later reference, we also state explicitly their derivatives,
	\ba
	F_1' &=& H_2'  - H_3 +3 \psi H_3' + 2 \psi^2 H_3'' + 4 H_4 - 
4 \psi H_4' + 20 \psi^2 H_4'' + 8 \psi^3 H_4''' \,,\label{eq:F1der}\\
	F_2' &=& - H_3' - 2 \psi H_3'' - 12 \psi H_4'' - 8 \psi^2 H_4'''\,,  \label{eq:F2der}\\
	F_3' &=& -H_3' - 4 \psi H_4''\,, \label{eq:F3der}\\
	F_4' &=&  H_3' + \psi H_3'' + 10\psi H_4'' + 4 \psi^2 H_4'''\,. \label{eq:F4der}
	\ea
To reconstruct the function $\Omega(\varphi,\psi)$, we observe that the second relation in~\eqref{eq:Omegarpsi} implies
	\ba\label{eq:OmegaPremature}
	\Omega(\varphi,\psi) &=& - \varphi^2 \int \dd{\psi} \beta(\varphi,\psi) + F(\varphi)\,.
	\ea
	Inserting this expression into the first relation in~\eqref{eq:Omegarpsi}, we obtain
	\ba
	\alpha(\varphi,\psi) &=&  - 2 \varphi \int \dd{\psi} \beta(\varphi,\psi) -\varphi^2\int\text{d}\psi\,\partial_\varphi\beta(\varphi,\psi) + 2 \varphi \psi \beta(\varphi,\psi) + F'(\varphi) \,.
	\ea
	Inserting the required form~\eqref{eq:beta} of $\beta(\varphi,\psi)$ into the previous equation leads to
	\ba
	\alpha(\varphi,\psi) &=& -  F_3(\psi) - 3\varphi^2 \int \dd{\psi} F_4(\psi) + 2 \psi F_3'(\psi) + 2 \varphi^2 \psi F_4(\psi)+F'(\varphi)\,\\
	&=& \varphi^2 \qty[- 3 \int \dd{\psi}F_4(\psi) + 2 \psi F_4(\psi)] - F_3(\psi) + 2 \psi F_3'(\psi)+F'(\varphi)\,.
	\ea
    Comparing with~\eqref{eq:alpha}, we see that it must be $F'(\varphi) = 0$. We proceed by setting $F(\varphi) \equiv 0$ without loss of generality. Thus we have to identify
    \ba
	F_1(\psi) &=& - 3 \int \dd{\psi}F_4(\psi) + 2 \psi F_4(\psi)\,,\label{eq:F1Need}\\
	F_2(\psi) &=&  -  F_3(\psi) + 2 \psi F_3'(\psi)\,. \label{eq:F2Need}
	\ea
	These conditions must be satisfied in order for $\alpha(\varphi,\psi)$ and $\beta(\varphi,\psi)$ computed from $\Omega(\varphi,\psi)$ with $F(\varphi)$ in~\eqref{eq:OmegaPremature} to satisfy the second equation of motion~\eqref{eq:Eq2} or, equivalently, equation~\eqref{eq:Eq2Rewritten}. As a cross check, one may verify that directly inserting $\alpha(\varphi,\psi)$ and $\beta(\varphi,\psi)$ as given in~\eqref{eq:alpha}--\eqref{eq:beta} into equation~\eqref{eq:Eq2Rewritten} leads to the two independent conditions
	\ba
	 F_1'(\psi)+F_4(\psi)-2\psi F_4'(\psi)&= & 0\,,\\
	F_2'(\psi)-F_3'(\psi)-2\psi F_3''(\psi)&= & 0\,,\label{eq:F2DerNeed}
	\ea
	which are just the differentiated versions of~\eqref{eq:F1Need}--\eqref{eq:F2Need}. The second equation is automatically satisfied by definition of $F_2(\psi)$ and $F_3(\psi)$ in~\eqref{eq:F2}--\eqref{eq:F3}, whereas the first equation imposes the following constraint among the function $H_i(\psi)$ in the Horndeski action~\eqref{eq:SHorndeskiSubclass},
    \ba
    H_2'(\psi)-H_3(\psi)+2\psi H_3'(\psi)+2H_4(\psi)- 2\psi H_4'(\psi)+4\psi^2H_4''(\psi)  & = & 0\,.\label{eq:HConstraint}
    \ea
    This relation must be satisfied for a given such
    theory to admit single-function static spherically symmetric solutions. It should be emphasized that this constraint applies specifically to the action written in the form~\eqref{eq:SHorndeskiSubclass}. Performing partial integrations in the action in general does not preserve the functional form $H_i = H_i(\psi)$ and the only invariant statement relevant to the existence of single-function static spherically symmetric solutions is the integrability condition
    \ba
    \partial_\chi  \alpha - \partial_\varphi \beta &=& 0\,,
\ea
where $\alpha$ and $\beta$ are expressed in terms of the variables $\varphi$ and $\chi$.\\

We will refer to non-polynomial gravities with second-order equations of motion on spherically symmetric backgrounds, which satisfy the integrability condition, as {\it non-polynomial quasi-topological gravities}. A constraint of the above form among the functions $H_i(\psi)$ is automatically generated in higher-dimensional polynomial gravities with second-order equations of motion on spherically symmetric backgrounds, which establishes the equivalence of such notion to polynomial quasi-topological gravities~\cite{Bueno:2025qjk}. We shall review the relevant statements later in Subsection~\ref{SecSub:ConnectionQTG}. Here we emphasize that the  constraint~\eqref{eq:HConstraint} singles out a strictly smaller subset of all the non-polynomial gravities with second-order equations of motion on spherically symmetric backgrounds.\\

Computing the function $\Omega(\varphi,\psi)$ in~\eqref{eq:OmegaPremature}, taking into account the constraints~\eqref{eq:F1Need}--\eqref{eq:F2Need}, leads to
\ba
\Omega(\varphi,\psi) &=& - \varphi \int \dd{\psi} F_3'(\psi) - \varphi^3 \int  \dd{\psi} F_4(\psi)\,,\\
&=&- \varphi F_3(\psi) + \frac{ \varphi^3}{3} \qty[F_1(\psi) - 2 \psi F_4(\psi)]\,,\\
&=& \varphi \qty[H_3(\psi) - 4 H_4(\psi) + 4 \psi H_4'(\psi)] + \frac{\varphi^3}{3} \qty[H_2(\psi) - \psi H_3(\psi) + 8 \psi H_4(\psi) - 8 \psi^2 H_4'(\psi)]\,\quad \quad \\
&\equiv &  \varphi G(\psi) + \frac{\varphi^3}{3} \qty[H(\psi) - 2\psi G(\psi) ]\label{eq:OmegaLong}\,,\quad \quad 
\ea
where we have defined
\ba
G(\psi) &=& H_3(\psi) - 4 H_4(\psi) + 4 \psi H_4'(\psi)\,,\\
H(\psi) &=& H_2(\psi) + \psi H_3(\psi)\,.
\ea
This allows us to compute explicitly the functions $\alpha(\varphi,\psi)$ and $\beta(\varphi,\psi)$ in terms of $G(\psi)$ and $H(\psi)$ for theories satisfying the integrability condition,
\ba
\alpha(\varphi,\psi) &=& \varphi^2 \qty[H - \frac{2}{3}\psi G - \frac{2}{3}\psi H' + \frac{4}{3}\psi^2 G'] + G - 2 \psi G'\,,\\
\beta(\varphi,\psi) &=& -\frac{1}{\varphi}G' + \frac{1}{3}\varphi \qty[-H' + 2 G + 2 \psi G']\,.
\ea
Herewith the remaining equation of motion~\eqref{eq:Eq1Again} with $\varphi = r$ and $\psi = (1-f)/r^2$ onshell can be integrated to yield the following algebraic equation determining the metric function $f$,
\ba \label{eq:EqMasterHG}
r (1+2 f)G\qty(\frac{1-f}{r^2}) + r^3 H\qty(\frac{1-f}{r^2}) &=& 12M\,,
\ea
where $M$ is an integration constant. This algebraic equation characterising static spherically symmetric solutions to four-dimensional non-polynomial quasi-topological gravities is one of the main results in this section. In the remainder of this work we will only analyse a subset of these theories for which $G\equiv 0$.\\
 
Concretely, we will proceed to consider an exemplary class of theories satisfying~\eqref{eq:HConstraint} by demanding separately the two identities
\ba
-2 H_2'(\psi) + H_3(\psi) - 2 \psi H_3'(\psi) &=& 0\,,\label{eq:H2H3}\\
H_3(\psi) - 2 \psi H_3'(\psi) - 4 H_4(\psi) + 4 \psi H_4'(\psi) - 8 \psi^2 H_4''(\psi) &=& 0\,.
\ea
The last one can be solved by setting
\ba
H_3(\psi) = 4 H_4(\psi) - 4 \psi H_4'(\psi) \,\,\, \quad \,\,\, \Leftrightarrow \,\,\, \quad \,\,\, G(\psi) \,\,= \,\,0 \,.\label{eq:H4H3}
\ea
The previous two differential equations can be solved explicitly in the form
	\ba
H_2(\psi) &=& 
\Lambda +
\frac{1}{2} \int \dd{\psi} \qty[H_3(\psi)  - 2 \psi H_3'(\psi)]\,,\label{eq:H2Sol}\\
H_4(\psi) &=& \Lambda' \psi 
- \frac{1}{4} \psi \int \dd{\psi} \frac{H_3(\psi)}{\psi^2}\,,\label{eq:H4Sol}
\ea
where $\Lambda$ and $\Lambda'$ are integration constants which we will set to zero in the following. By choosing $H_2(\psi)$ and $H_4(\psi)$ as above, the second equation of motion~\eqref{eq:Eq2Rewritten} is guaranteed to be satisfied. In particular, the theories satisfying the constraints~\eqref{eq:H2H3} and~\eqref{eq:H4H3} are parametrized by a single one-variable function of $\psi$.\\

For illustrative purposes, in the following we will repeat the steps leading to the algebraic equation determining $f(r)$ for static spherically symmetric solutions to the special class of non-polynomial quasi-topological gravities satisfying $G(\psi) = 0$. As a result of the above restrictions, the functions $F_i(\psi)$ take the form
	\ba
	F_1(\psi) \,\,=\,\,H_2(\psi) \,,\,\,\,\quad 
	F_2(\psi) \,\,=\,\, 0\,,\,\,\,\quad 
	F_3(\psi) \,\,=\,\,0\,,\,\,\,\quad 
	F_4(\psi) \,\,=\,\, - \frac{1}{2}H_3(\psi)\,.
	\ea
	Therefrom we conclude that the functions $\alpha(\varphi,\psi)$ and $\beta(\varphi,\psi)$ for these theories take the form
	\ba\label{eq:AlphaBetaConstrained}
	\alpha(\varphi,\psi) \,\, = \,\, \varphi^2 H_2(\psi)\,\,\, \quad \text{and} \quad \,\,\, \beta(\varphi,\psi)\,\, =\,\, -\frac{1}{2} \varphi H_3(\psi)\,.
	\ea
	We can now state the function
	$\Omega(\varphi,\psi)$ 
	for these theories. 
This function has to satisfy the relations
	\ba \label{eq:omegaint}
	\partial_\varphi \Omega(\varphi,\psi) &=& \varphi^2 \qty[H_2(\psi) + \psi H_3(\psi)]\,\,\,\quad \text{and} \quad \,\,\,\partial_\psi \Omega(\varphi,\psi) \,\,=\,\, \frac{1}{2}\varphi^3 H_3(\psi)\,,
	\ea
	whose solution is given by
	\ba\label{eq:omegaSol}
	\Omega(\varphi,\psi) &=& \frac{1}{3}\varphi^3 H(\psi) + \Omega_0 \,\,\,\quad \text{where} \quad \,\,\, H(\psi) \,\,\equiv \,\,H_2(\psi) + \psi H_3(\psi)\,,
	\ea
	provided that~\eqref{eq:H2H3} is satisfied.\\
	
In the following, it will be convenient to write all expressions in terms of the function
\ba
H(\psi) &=& H_2(\psi) + \psi H_3(\psi)\,,\label{eq:EqH}\\
H'(\psi) &=& H_2'(\psi) + H_3(\psi )+ \psi H_3'(\psi) \,\,= \,\, \frac{3}{2}H_3(\psi)\,.\label{eq:EqHp}
\ea
The remaining equation of motion~\eqref{eq:Eq1} then becomes
\ba
r^2 H_2(\psi) - \frac{1}{2} r H_3(\psi) f' \,\, = \,\, 0 \,\,\,\quad \,\,\, \Leftrightarrow  \,\,\,\quad \,\,\, 
H(\psi) - \frac{1}{3}\qty(2  \psi  +   \frac{f'}{r})H'(\psi)\,\,=\,\, 0
\ea
with $ \psi(r,f) =  (1-f)/r^2$,
which can be integrated to yield
\ba\label{eq:EqMaster}
H\qty(\frac{1-f}{r^2}) \,\, =\,\, \frac{12 M}{r^{3}}\,,
\ea
where $M$ is an integration constant related to the ADM mass of the solution.~\footnote{As a cross-check, inserting $H_2(\psi) = 2\psi$ and $H_3(\psi)=4 $ as in general relativity, results in $H(\psi) = 6 \psi$ and leads to the Schwarzschild spacetime with function $f(r) = 1 -2M/r$.} As a result, the 
the function $\Omega(\varphi,\psi)$ onshell is proportional to $M$, i.e,
\ba\label{eq:omegaM}
\Omega\qty(r,\,\frac{1-f}{r^2}) &=& 4  M\,.
\ea 
In particular, as anticipated, the function $H(\psi)$ defined in~\eqref{eq:omegaSol} determines the metric function $f(r)$ through the algebraic equation~\eqref{eq:EqMaster}. 

\subsection{Rederivation of the equations of motion for subclasses of non-polynomial quasi-topological gravities}\label{SecSub:EOMnf}

In this subsection, for completeness, we will explicitly rederive the equations of motion on static spherically symmetric backgrounds for the particular class of non-polynomial gravitational theories~\eqref{eq:SLift} parametrized in terms of a single-variable function $H(\psi)$ as discussed at the end of the previous subsection, which form a subclass of  non-polynomial quasi-topological gravities.  Our goal is to reveal the Lagrangian mechanism that ensures solutions to depend on a single metric function $f(r)$ which satisfies an algebraic equation.  This will also serve later to make the connection between these theories and polynomial quasi-topological gravities~\cite{Bueno:2025qjk}. At the same time, the discussion here illustrates that the solution space of the more general class of  non-polynomial gravities with reduced actions~\eqref{eq:SHorndeskiSubclass}, for which the functions $H_i(\psi)$ are not constrained in the way~\eqref{eq:HConstraint}, and hence which are not of non-polynomial quasi-topological type, is larger than the space of single-function spacetimes considered in this section. 

Concretely, we start from the non-polynomial gravitational actions constructed in~\eqref{eq:SLift}, i.e.,
\ba\label{eq:SLiftAgain}
S[g] &=& \int \dd[4]{x} \sqrt{-g}\, \qty[H_2(\mathcal{P}) - H_3(\mathcal{P}) \mathcal{H} + H_4(\mathcal{P})\mathcal{K} + 2  H_4'(\mathcal{P}) \qty(\mathcal{H}^2 - \mathcal{T}) ]\,,
\ea
and evaluated on generic static spherically symmetric spacetimes
\ba\label{eq:MetricSSS}
g_{\mu\nu}(x)\dd{x}^\mu \dd{x}^\nu &=& -n(r)^2f(r)\dd{t}^2 + \frac{\dd{r}^2}{f(r)} + r^2 \dd{\Omega}^2\,.
\ea
For this class of spacetimes, the four combinations defined in~\eqref{eq:BuildingBlocks} are explicitly given by
\ba
\eval{ \qty{\mathcal{R}\,,\, \eta\,,\, \tau\,,\,\psi}}_{n,f} & = & \bigg\{- f'' - \frac{1}{n}\qty[2 f n'' + 3 f' n']\,,\, \frac{1}{rn}\qty[n f' + f n'] \,,\, \label{eq:BuildingBlocksnf}\\
&{}& \quad \quad 	\frac{1}{2 r^2 n^2}\qty[f'^2 n^2+ 2 f f' n n' + 2 f^2 n'^2]\,,\, \frac{1-f}{r^2}\bigg\}\,.
\ea
Disregarding a factor $\Omega_2$ and the constant $t$-integral, the reduced action is thus given by
\ba
S[n,f] &\equiv & \int \dd{r} L_{n,f} \,\,= \,\, \int \dd{r} n r^2 \mathcal{L}_{n,f}\\
&=&  \int \dd{r} n r^2 \, \eval{\qty[H_2(\psi) - H_3(\mathcal{P}) \eta + H_4(\psi)\mathcal{R} + 2  H_4'(\psi) \qty(\eta^2 - \tau) ]}_{n,f}
\ea
with
\ba
\mathcal{L}_{n,f}&=& H_2(\psi) - \frac{1}{rn} \qty[n f' + f n'] H_3(\psi)  + \frac{1}{n}\qty[-n f'' - 2 f n'' - 3 f' n'] H_4(\psi)   \nn\\
&-& \frac{1}{n r^2} \qty[n f'^2 + 2 f f'n']H_4'(\psi)\,,
\ea
where $\psi(r,f) = (1-f)/r^2$. For future reference, we also introduce the notation $\mathcal{L}_f \equiv \eval{\mathcal{L}_{n,f}}_{n=1}$, which represents the reduced Lagrangian density $\mathcal{L}_{n,f}$ evaluated for $n=1$.\\

As explained above, we will consider specifically the restricted actions of non-polynomial quasi-topological type with functions $H_{i}(\psi)$ chosen according to~\eqref{eq:H4Sol}, \eqref{eq:H2Sol} and~\eqref{eq:omegaSol}, i.e., determined by a single-variable function $H(\psi)$.  In this case, it is useful to rewrite the above expression in terms of 
\ba
H(\psi) &=& H_2(\psi) + \psi H_3(\psi)\,,\label{eq:H}\\
H'(\psi) &=& H_2'(\psi) + H_3(\psi) + \psi H_3'(\psi) \,\,= \,\, \frac{3}{2}H_3(\psi) \label{eq:Hp}\,,
\ea
cf.~equations~\eqref{eq:EqH} and~\eqref{eq:EqHp}, such that
\ba\label{eq:HiFuncH}
H_3(\psi) \,\,=\,\, \frac{2}{3} H'(\psi)\,, \quad \quad H_2(\psi) \,\,=\,\, H(\psi) - \frac{2}{3}\psi H'(\psi)\,,\quad H_4(\psi) \,\,=\,\, -\frac{1}{6}\psi \int \dd{\psi} \frac{H'(\psi)}{\psi^2}\,.
\ea
The result can be expressed as
\ba\label{eq:LnfRewritten}
L_{n,f} &=& n r^2 \mathcal{L}_{n,f} \,\,=\,\, \derivative{}{r}\qty[\frac{1-f}{6}\qty(n f' + 2 fn') \int \dd{\psi} \frac{H'(\psi)}{\psi^2}] + n \derivative{}{r}\qty[\frac{1}{3}r^3 H\qty(\frac{1-f}{r^2})]\,.
\ea

From~\eqref{eq:LnfRewritten} is it straightforward to obtain the Euler-Lagrange equations for $n$ and $f$. The first term is a total derivative and hence does not affect the equations of motion. The variation of the second term leads to
\ba\label{eq:EqMasternH0}
H'(\psi) n' \,\,=\,\, 0 \,\,\, \quad \text{and} \quad \,\,\, \derivative{}{r}\qty[r^3 H(\psi)] \,\,=0\,.
\ea
Thus, the unique  static spherically symmetric solutions to the non-polynomial four-dimensional gravities with functions $H_i(\psi)$ related as above are characterized by
\ba\label{eq:EqMasternH}
n' \,\,=\,\, 0  \quad \text{and} \quad \,\,\, H(\psi) \,\, =\,\, \frac{12 M}{r^3}\,,
\ea
where $M$ is an integration constant proportional to the ADM mass of the spacetime once a specific choice for the theory-dependent function $H(\psi)$ has been made. The relations~\eqref{eq:EqMasternH0} and~\eqref{eq:EqMasternH} are the central result in this section and show that static spherically symmetric solutions to the one-function non-polynomial quasi-topological gravities specified above are described by a single metric function satisfying an algebraic equation involving a single one-variable function of $\psi(r,f)$.\\

It is important to emphasize that the solution space to the above algebraic equation is larger than the solution space of the restricted class of non-polynomial quasi-topological gravities constructed in~\cite{Bueno:2025zaj}. This extension arises from the fact that the theories considered in~\cite{Bueno:2025zaj} can only produce two-dimensional Horndeski actions in which $H(\psi)$ is a polynomial function at each finite effective order in the curvature. In particular, resummations of an infinite tower of therein provided non-polynomial quasi-topological densities are required to generate non-analytic characteristic functions $H(\psi)$ and thereby generate some of the regular black-hole spacetimes that have been found as solutions to higher-dimensional quasi-topological gravities~\cite{Bueno:2024dgm}, as solutions to four-dimensional gravities. Not all four-dimensional analogues of the spacetimes generated in~\cite{Bueno:2024dgm} can be obtained from the construction in~\cite{Bueno:2025zaj}, because this construction skips contributions from certain orders in the curvature, such at order two where the Gauss-Bonnet invariant is used as a density. As an example, with this gap it is not possible to build the resummation series needed to generate the Hayward spacetime as a solution. Such obstructions do not arise in the construction presented here, as by lifting the quantity $\psi$ in the Horndeski action independently to a four-dimensional covariant density, we are able to produce arbitrary non-analytic functions $H(\psi)$ in the above equation.

\subsection{Remark on generic static spherically symmetric solutions}

In this subsection,  we emphasize for completeness that the solution space of the Horndeski actions~\eqref{eq:SHorndeskiSubclass} which can be generated from the reduction of non-polynomial gravities such as~\eqref{eq:SLift} is genuinely larger than the one of non-polynomial quasi-topological gravities. For the general static spherically symmetric ansatz~\eqref{eq:MetricSSS}, the two independent equations of motion~\eqref{eq:FieldEq} are given by
\ba
\mathbb{G}_{tt} &=&
\frac{1}{r^2}\mathcal{E}_{tt}  \,\, = \,\, 
\frac{n^2 f}{2 r^2}\qty[\alpha + \beta f'] \,\, =\,\,0 \,,\label{eq:EOMttnf}\\
\mathbb{G}_{rr} &=& \frac{1}{r^2}\mathcal{E}_{rr} 
\,\,= \,\,-\frac{1}{2  r^2n f}\qty[n\qty(\alpha + \beta f') + 2 fn' \beta - 2 n f\qty(\partial_f \alpha - \partial_r \beta) ] \,\, =\,\, 0\,.\label{eq:EOMrrnf}
\ea
If a non-polynomial gravity with second-order equations of motion on~\eqref{eq:MetricSSS} does not belong to the quasi-topological class, then it will be
\ba\label{eq:Integrability}
\partial_f \alpha - \partial_r \beta &=& - H_2' + H_3 - 2 \psi H_3' - 2 H_4 + 2 \psi H_4' - 4 \psi^2 H_4'' \,\, \neq 0\,,
\ea
i.e., the integrability constraint~\eqref{eq:HConstraint} among the $H_i$ is not satisfied.
In this case, for any given theory-dependendent choice of functions $\alpha(\varphi,\chi)$ and $\beta(\varphi,\chi)$, with $\varphi = r$ and $\chi = f$ onshell, the first equation~\eqref{eq:EOMttnf} determines the function $f(r)$, and in turn the second equation~\eqref{eq:EOMrrnf} determines the function $n(r)$,
\ba\label{eq:EOMnf}
f' \,\,=\,\, -\frac{\alpha}{\beta}\,\,\, \quad \text{and} \quad \,\,\, \frac{n'}{n} \,\,=\,\, \frac{\partial_f \alpha - \partial_r \beta}{\beta}\,,
\ea
see also~\cite{Borissova:2026dlz}, where we assume that $\beta$ is non-zero. In particular, solutions in this case are generically spacetimes with two distinct functions in their line element.\\

\subsection{Connection to polynomial quasi-topological gravities}\label{SecSub:ConnectionQTG}

In this subsection, we will establish connections of the previous results to statements on polynomial quasi-topological gravities derived recently in~\cite{Bueno:2025qjk}, whereby the action~\eqref{eq:S} is assumed to be a polynomial function of curvature invariants. Here, the assumption about polynomiality  does not apply as our construction yields gravitational actions that are manifestly non-polynomial in curvature invariants. However, we find that some statements derived in~\cite{Bueno:2025qjk} apply also to the non-polynomial quasi-topological gravities considered in previous subsections,  as  the corresponding proofs rely on the form of the spherically reduced actions, which in all cases are Horndeski actions of the type considered here. On the other hand, we will also see that the majority of statements applicable to polynomial gravities with second-order equations of motion on spherically symmetric spacetimes do not, in general, apply to non-polynomial gravities with second-order equations of motion on spherically symmetric spacetimes, unless these are of quasi-topological type and parametrized in terms of the previously defined single function $H$.
\\

To begin, using~\eqref{eq:BuildingBlocksnf} we note that the evaluation of the special  subclass of Horndeski Lagrangian densities~\eqref{eq:SHorndeskiSubclass} for the single-function configurations~\eqref{eq:AnsatzSSSn1} is given by
\ba\label{eq:LfHorndeski}
{\mathcal{L}_{\text{Horndeski}}}_f &=& H_2(\psi) - H_3(\psi) \frac{f'}{r} - H_4(\psi)f'' +  H_4'(\psi)\frac{f'^2}{r^2}\,,
\ea
where $\psi = \psi(r,f)$. Since it is these classes of two-dimensional Horndeski actions which we have constructed as the reduced actions of four-dimensional non-polynomial gravities, by construction, for any of the resulting generally covariant four-dimensional Lagrangians it holds $\mathcal{L}_f = {\mathcal{L}_{\text{Horndeski}}}_f $, which can be written as
\ba\label{eq:Lf}
\mathcal{L}_f &=&  H_2(\psi) -H_3(\psi) \frac{f'}{r} - H_4(\psi)f'' -  \qty(\partial_f H_4(\psi))f'^2\,,
\ea
where we remind that $\mathcal{L}_f$ is defined as the reduced Lagrangian density on~\eqref{eq:MetricSSS}, $\mathcal{L}_{n,f}$, evaluated for $n=1$.
The analogous statement extended to generic static spherically symmetric configurations with line element~\eqref{eq:MetricSSS} is
\ba\label{eq:LnfHorndeski}
{\mathcal{L}_\text{Horndeski}}_{\,n,f}&=& H_2(\psi) - \frac{1}{rn} \qty[n f' + f n'] H_3(\psi)  + \frac{1}{n}\qty[-n f'' - 2 f n'' - 3 f' n'] H_4(\psi)   \nn\\
&-& \frac{1}{n r^2} \qty[n f'^2 + 2 f f'n']H_4'(\psi)\,,
\ea
and the reduced Lagrangian density is $\mathcal{L}_{n,f} = {\mathcal{L}_\text{Horndeski}}_{\,n,f}$, which can be written as
\ba\label{eq:Lnf}
\mathcal{L}_{n,f}&=& -H_4(\psi)f'' - \qty(\partial_f H_4(\psi)) f'^2 - H_3(\psi) \frac{f'}{r} + H_2(\psi)\nn\\
&+& \frac{f'n'}{n}\qty[-3 H_4(\psi) - 2 f  \qty(\partial_f H_4(\psi))] - \frac{f n'}{rn} H_3(\psi) - \frac{2f}{n}H_4(\psi) n''\,.
\ea

\noindent This shows that the non-polynomial gravities constructed here satisfy (in adapted notation):
\\

 {\bf Proposition 1} in~\cite{Bueno:2025qjk}: {\it Let $\mathcal{L}(g^{\mu\nu}, R_{\alpha\beta\gamma\delta})$ be a higher-curvature gravity  built from polynomial combinations of curvature invariants. Assume its equations of motion on~\eqref{eq:MetricSSSn1} contain at most second derivatives of $f(r)$. Then, the evaluation of $\mathcal{L}(g^{\mu\nu}, R_{\alpha\beta\gamma\delta})$ on the background~\eqref{eq:MetricSSSn1}, denoted as $\mathcal{L}_f$, takes the form~\eqref{eq:Lf} for certain single-variable functions $H_2$, $H_3$ and $H_4$.}
\\

 {\bf Proposition 5} in~\cite{Bueno:2025qjk}: {\it Let $\mathcal{L}(g^{\mu\nu}, R_{\alpha\beta\gamma\delta})$ be a higher-curvature gravity  built from polynomial curvature invariants with second-order equations on a generic static and spherically symmetric ansatz~\eqref{eq:MetricSSS}. Then $\mathcal{L}_{n,f}$ reads as follows: \eqref{eq:Lnf}.}\\

\noindent  --- and more generally they satisfy the generalizations of these propositions to dynamical spherically symmetric spacetimes with generic functions $f(t,r)$ and $n(t,r)$ as formulated in {\bf Proposition 8} in~\cite{Bueno:2025qjk}. The assumption about polynomiality of the action in curvature invariants can be relaxed, as any gravitational theory built (not necessarily polynomially) from polynomial Riemann invariants that yields up to second-order equations of motion on generic (and hence on subclasses of) warped-product backgrounds~\eqref{eq:Metric}, by construction and in accordance with the principle of symmetric criticality, must yield a two-dimensional Horndeski action of the more general form~\eqref{eq:SHorndeskiSubclass}. However, it is important to emphasise that in the non-polynomial case, $\mathcal{L}_f$ must be computed explicitly as the evaluation of the reduced Lagrangian density $\mathcal{L}_{n,f}$ for $n=1$, i.e., {\it not} as the immediate evaluation of $\mathcal{L}$ on~\eqref{eq:MetricSSSn1}. This is because the covariant densities to which the Horndeski elements are lifted may not be well-defined on the restricted choice of single-function static ansatz~\eqref{eq:MetricSSSn1}, but are well-defined on a generic static spherically symmetric ansatz which must be adopted in any case to comply with the variational principle. The same remark applies, e.g., also to the non-polynomial gravities constructed in~\cite{Bueno:2025zaj}.
\\

On the other hand, we see that the following proposition derived for polynomial quasi-topological gravities~\cite{Bueno:2025qjk} can only be applied to the restricted subclass of  non-polynomial quasi-topological gravities specified at the end of Subsection~\ref{SecSub:HorndeskiOneFunction} and in considered in detail in Subsection~\ref{SecSub:EOMnf}, and only in an extended sense.\\

\noindent 
Consider e.g.~(in adapted notation and $D=4$):
\\

	{\bf Proposition 2} in~\cite{Bueno:2025qjk}: {\it Let $\mathcal{L}(g^{\mu\nu}, R_{\alpha\beta\gamma\delta})$ be a higher-curvature gravity  built from polynomial combinations of curvature invariants. Assume its equations of motion on~\eqref{eq:MetricSSSn1}  are of second order in derivatives. The functions $H_2$, $H_3$ and $H_4$ in~\eqref{eq:Lf} can be entirely parametrized as follows:}
\ba\label{eq:HP2}
H_4(\psi) \,\, =\,\, - \frac{1}{2} \psi \int \dd{\psi} \frac{h'(\psi)}{\psi^2}\,,\quad \quad  H_3(\psi) \,\,=\,\, 2 h'(\psi) \,,\quad \quad H_2(\psi) \,\,=\,\, 3 h(\psi) - 2 \psi h'(\psi)\,,\quad \quad 
\ea
{\it where $h$ is an arbitrary analytic function of $\psi$.}
\\

\noindent 
In fact, the above identification of functions $H_i(\psi)$ can be rewritten as
\ba
H_4(\psi) & = & - \frac{1}{4} \psi \int \dd{\psi} \frac{H_3(\psi)}{\psi^2}\,,\quad \quad H_2(\psi) + \psi H_3(\psi) \,\, =\,\, 3 h(\psi)\,,\\
 H_2(\psi) &=& \frac{3}{2}\int \dd{\psi} \qty[H_3(\psi)] - \psi H_3(\psi)\,,
\ea
which is in accordance with the result~\eqref{eq:H4Sol} for $H_4(\psi)$ and~\eqref{eq:omegaSol} with $H(\psi) =3 h(\psi)$, as well as~\eqref{eq:H2Sol} for $H_2(\psi)$, rewritten in the form
\ba\label{eq:H2SolRewritten}
H_2(\psi) &=& 
\frac{1}{2} \int \dd{\psi} \qty[H_3(\psi)  - 2 \psi H_3'(\psi)] \,\,=\,\, \frac{1}{2} \int \dd{\psi} \qty[3 H_3(\psi)] - \psi H_3(\psi)
\ea
using partial integration. However, $H(\psi)$ in our case does not need to be an analytic function of $\psi$. Moreover, it is clear that the four-dimensional non-polynomial gravities constructed in Subsection~\ref{SecSub:Expressing2DElementsAs4DCovariantDensities} with reduced actions~\eqref{eq:SHorndeskiSubclass}, will have second-order equations of motion on~\eqref{eq:MetricSSSn1} (as they do on generic warped-product spacetimes~\eqref{eq:Metric}), but they will not, in general, feature  functions $H_i(\psi)$ that can be parametrized in terms of one function $h(\psi)$ as above, unless they are of special non-polynomial quasi-topological type such as  the ones considered at the end of Subsection~\ref{SecSub:HorndeskiOneFunction} and in Subsection~\ref{SecSub:EOMnf}. In fact, let us remind that here we have only analyzed a specific subclass of non-polynomial quasi-topological gravities for which the constraint among the $H_i(\psi)$~\eqref{eq:HConstraint}, necessary for the existence of single-function static spherically symmetric solutions, was satisfied my imposing separately~\eqref{eq:H2H3} and~\eqref{eq:H4H3}. In general, this restriction is not strictly necessary and illustrates that the actual space of reduced non-polynomial quasi-topological gravities is larger, cf.~the general discussion in Subsection~\ref{SecSub:HorndeskiOneFunction} involving the two functions $H(\psi)$ and $G(\psi)$.

In polynomial quasi-topological gravities, an analogue constraint~\eqref{eq:HConstraint} relating the functions $H_i(\psi)$ arises from the sole condition that the reduced action must originate from an actual polynomial combination of curvature invariants. In particular, in this case these functions automatically satisfy the separate conditions~\eqref{eq:H2H3} and~\eqref{eq:H4H3} and are thereby parametrizable in terms of one (analytic) function $H(\psi)$, and in particular satisfy $G(\psi) =0$. By contrast, the construction of the non-polynomial actions~\eqref{eq:SLift} relies on lifting the elements in the two-dimensional Horndeski actions~\eqref{eq:SHorndeskiSubclass} independently, such that no such constraint arises a priori and needs to be imposed in addition, in order to obtain single-function static spherically symmetric solutions characterized by~\eqref{eq:EqMaster} rather than~\eqref{eq:EqMasterHG} for a generic function $G(\psi)$. 

To give an example, the proof of the above proposition for polynomial quasi-topological gravities in~\cite{Bueno:2025qjk} makes use of a Ricci--Weyl decomposition of curvature invariants used to polynomially construct the Lagrangian density $\mathcal{L}\qty(g^{\mu\nu},R_{\alpha\beta\gamma\delta})$, and a rewriting of the elements appearing in the reduced Lagrangian density $\mathcal{L}_f$ as follows,~\footnote{Recall that for spacetimes~\eqref{eq:MetricSSS}, there are only three independent invariants parametrizing the Riemann tensor, cf.~\eqref{eq:BuildingBlocksnf} with $n=1$.}
\ba
\eval{\qty{\mathcal{R}\,,\, \eta\,,\, \psi}}_{f} \,\,\equiv \,\, \qty{- f''\,,\, \frac{f'}{r} \, ,\,\frac{1-f}{r^2}} \,\,\,\quad \leftrightarrow \,\,\,\quad \qty{\Omega_f\,,\, \Theta_f \,,\, \rho_f}\,,
\ea
where $\Omega_f$, $\Theta_f$ and $\rho_f$ are the scalar expressions parametrizing the Weyl tensor, the traceless Ricci tensor and the Ricci scalar for  the spacetimes~\eqref{eq:MetricSSSn1}, i.e., up to proprotionality factors,
\ba
\Omega_f &=& \frac{2 - 2 f + 2 r f' - r^2 f''}{r^2}\,,\\
\Theta_f &=& \frac{2 - 2f + r^2 f''}{r^2}\,,\\
\rho_f &=& \frac{2 - 2 f - 4 r f' - r^2 f''}{r^2}\,,
\ea
as can be verified by making use of equations~\eqref{eq:RicciTensor}--\eqref{eq:Omega}. 
The fact that the traceless Ricci tensor is diagonal is a specific property of the metrics~\eqref{eq:MetricSSSn1}. The above translation $\eval{\qty{\mathcal{R} , \eta, \psi}}_f \,\leftrightarrow \, \qty{\Omega_f, \Theta_f, \rho_f}$ is in particular provided by a linear map. If $\mathcal{L}(g^{\mu\nu},R_{\alpha\beta\gamma\delta})$ is constructed polynomially from curvature invariants, then $\mathcal{L}_f$ will be a power series in $\Omega_f$, $\Theta_f$ and $\rho_f$ which has to satisfy certain constraints. For example, no terms linear in $\Omega_f$ and $\Theta_f$ with a purely $\rho_f$-dependent prefactor, can appear, as the Weyl and traceless Ricci tensors need to be contracted into a scalar. Such a constraint does not arise if non-polynomial combinations of curvature invariants are allowed. For example, we have seen (and explicitly made use of) that $\Omega$, defined for generic warped-product spacetimes in~\eqref{eq:Omega}, can be expressed covariantly as a ratio of subsequent Weyl invariants, cf., e.g., \eqref{eq:OmegaLift}. Similarly, $\Theta_f$ can be generated covariantly as a ration of subsequent traceless Ricci invariants evaluated on~\eqref{eq:MetricSSSn1}. Thus, in non-polynomial gravities with second-order equations on~\eqref{eq:MetricSSSn1}, the reduced Lagrangian density on~\eqref{eq:MetricSSSn1}, $\mathcal{L}_f$, may contain linear terms in $\Omega_f$ and $\Theta_f$ with arbitrary $\rho_f$-dependent prefactors. This makes the proof of the above proposition in~\cite{Bueno:2025qjk} inapplicable.\\

\noindent Consider (the relevant parts of, and in adapted notation, with $D=4$):\\

{\bf Proposition 3} in~\cite{Bueno:2025qjk}: {\it Let $\mathcal{L}(g^{\mu\nu}, R_{\alpha\beta\gamma\delta})$ be a higher-curvature theory built from polynomial contractions of curvature tensors. Then the following are equivalent:

\begin{itemize}
\item[2.] $\mathcal{L}(g^{\mu\nu}, R_{\alpha\beta\gamma\delta})$ possesses second-order equations of motion for the single-function static and spherically symmetric ansatz~\eqref{eq:MetricSSSn1}.
\item[3.] The Lagrangian $\mathcal{L}_f$ evaluated on~\eqref{eq:MetricSSSn1} is given by:
\ba
\mathcal{L}_{f} \,\,=\,\, \frac{1}{r^2}\derivative{}{r}\qty[\frac{(1-f)f'}{2} \int^\psi \dd{\kappa} \frac{h'(\kappa)}{\kappa^2} + r^3 h\qty(\frac{1-f}{r^2})]\,,\label{eq:LfP3}
\ea
where $r^2 \psi = 1-f$ and $h$ is an analytic function.
\end{itemize}
}
\hspace{1cm}

{\bf Proposition 6} in~\cite{Bueno:2025qjk}: {\it Let $\mathcal{L}(g^{\mu\nu}, R_{\alpha\beta\gamma\delta})$ be a higher-curvature theory built from arbitrary polynomial contractions of curvature tensors. The following are equivalent:
	
	\begin{itemize}
		\item[2.] The equations of motion for general static spherically symmetric configurations~\eqref{eq:Metric} are second order.
		\item[3.] The Lagrangian evaluated on~\eqref{eq:MetricSSS}, $\mathcal{L}_{n,f}$, reads:
		\ba
\mathcal{L}_{n,f} &=&  \frac{1}{n r^2}	\derivative{}{r}\qty[\frac{1-f}{2}\qty(n f' + 2 fn') \int^\psi \dd{\kappa} \frac{h'(\kappa)}{\kappa^2}] + \frac{1}{r^2} \derivative{}{r}\qty[r^3 h\qty(\frac{1-f}{r^2})]\,,\label{eq:LnfP6}
	\ea
		where  $h$ is an analytic function of its argument.
	\end{itemize}
}

\noindent 

Comparing the above expressions for $\mathcal{L}_{n,f}$ and $\mathcal{L}_f= \eval{\mathcal{L}_{n,f}}_{n=1}$ with~\eqref{eq:LnfRewritten}, we see that analogue statements are satisfied for the non-polynomial quasi-topological gravitational actions considered at the end of Subsection~\ref{SecSub:HorndeskiOneFunction} and in Subsection~\ref{SecSub:EOMnf} with the identification $H(\psi)=3h(\psi)$, which however does not need to be an analytic function of $\psi$. However, e.g., clearly, the more general class of non-polynomial gravitational actions~\eqref{eq:SLift} will lead to reduced Lagrangians of the form~\eqref{eq:LfHorndeski} and~\eqref{eq:Lnf} when evaluated on~\eqref{eq:MetricSSS} and~\eqref{eq:MetricSSS}, respectively, which cannot be written in the form~\eqref{eq:LfP3} and~\eqref{eq:LnfP6}, as their functions $H_i(\psi)$ need not be related to each other as asserted in~\eqref{eq:HP2}. In this case, one has to resort to the weaker statements of {\bf Proposition 1} and {\bf Proposition 5}.\\

\noindent 
Moreover, consider (in adapted notation and $D=4$):
\\

	{\bf Proposition 4} in~\cite{Bueno:2025qjk}: {\it Let $\mathcal{L}(g^{\mu\nu}, R_{\alpha\beta\gamma\delta})$ be a higher-curvature theory constructed from polynomial contractions of curvature tensors. If the equations of motion on~\eqref{eq:MetricSSSn1} feature no derivatives of $f$ of degree higher than two, then the unique single-function solutions~\eqref{eq:MetricSSSn1} are determined by a first-order differential equation for $f(r)$ which can be exactly integrated to yield:}
	\ba\label{eq:hEq}
	h\qty(\frac{1-f}{r^2}) &=& \frac{2M}{r^3}\,,
	\ea
	{\it where $M$ is an integration constant related to the ADM mass of the solution and $h$ is a theory-dependent function which may be computed as}
	\ba\label{eq:hEqGenerate}
	3 h\qty(\frac{1-f}{r^2}) &=& \eval{\mathcal{L}_f}_{f'\to 0, f''\to 0} - r \,\psi(r,f) \eval{\qty(\partial_{f'}\mathcal{L}_f)}_{f'\to 0}\,,
	\ea
	{\it where $f$, $f'$ and $f''$ are here conceived as if they were independent variables --- i.e., sending $f' \to0$ or $f'' \to 0$ does not imply anything on $f$ in the matter of this computation.}
\\

\noindent
For the special subclass of non-polynomial quasi-topological gravities considered at the end of Subsection~\ref{SecSub:HorndeskiOneFunction} and in Subsection~\ref{SecSub:EOMnf}, $\mathcal{L}_f$ is given by~\eqref{eq:Lf} with functions $H_i(\psi)$ that can be parametrized as~\eqref{eq:HiFuncH}, such that the right-hand side of the above equation evaluates to
\ba
H_2\qty(\frac{1-f}{r^2}) +  \qty(\frac{1-f}{r^2}) H_3\qty(\frac{1-f}{r^2}) &=& H\qty(\frac{1-f}{r^2})\,,
\ea
where we have inserted the definition of $H(\psi)$ in~\eqref{eq:omegaSol}, cf.~also~\eqref{eq:H}. Thus, as before, we identify $H(\psi) = 3 h(\psi)$ (not necessarily analytic) and, as we have seen, this function indeed determines $f$ by an algebraic equation of the form analogous to~\eqref{eq:hEq}, concretely equation~\eqref{eq:EqMaster} (cf.~also~\eqref{eq:EqMasternH}). Therefrom we conclude that the analogue statement of the above proposition is satisfied for this particular subclass of non-polynomial quasi-topological gravities. It does, however, not extend  to the generic class of non-polynomial gravities with reduced actions~\eqref{eq:SHorndeskiSubclass}, as for these theories to admit single-function solutions it is sufficient to demand only~\eqref{eq:HConstraint}, which results in the generalized algebraic equation~\eqref{eq:EqMasterHG}.\\
	
\noindent 
Finally, consider (in adapted notation and $D=4$):
\\

	{\bf Proposition 7} in~\cite{Bueno:2025qjk}: {\it The unique static spherically symmetric solutions~\eqref{eq:MetricSSS} of a theory $\mathcal{L}(g^{\mu\nu},R_{\alpha\beta\gamma\delta})$ built from polynomial contractions of curvature tensors and having second-order equations on~\eqref{eq:MetricSSS} are given by:}
\ba
h'(\psi) n' \,\,=\,\,0 \,, \quad \quad \derivative{}{r}\qty[r^3 h(\psi)] \,\,=\,\,0\,,
\ea
{\it where $h$ is a theory-dependent function given by~\eqref{eq:hEqGenerate} in Proposition 4.}
\\

\noindent This proposition does not extend to the general class of non-polynomial gravities with reduced actions~\eqref{eq:SHorndeskiSubclass}, which have second-order equations on~\eqref{eq:MetricSSS}, but for which static spherically symmetric solutions feature in general two free functions determined by~\eqref{eq:EOMnf} --- unless the theory is a non-polynomial quasi-topological gravity of the particular subtype discussed previously (and in this case $h(\psi)$ does not need to be analytic).\\

\begin{figure}[t]
	\centering
	\includegraphics[width=1.0\textwidth]{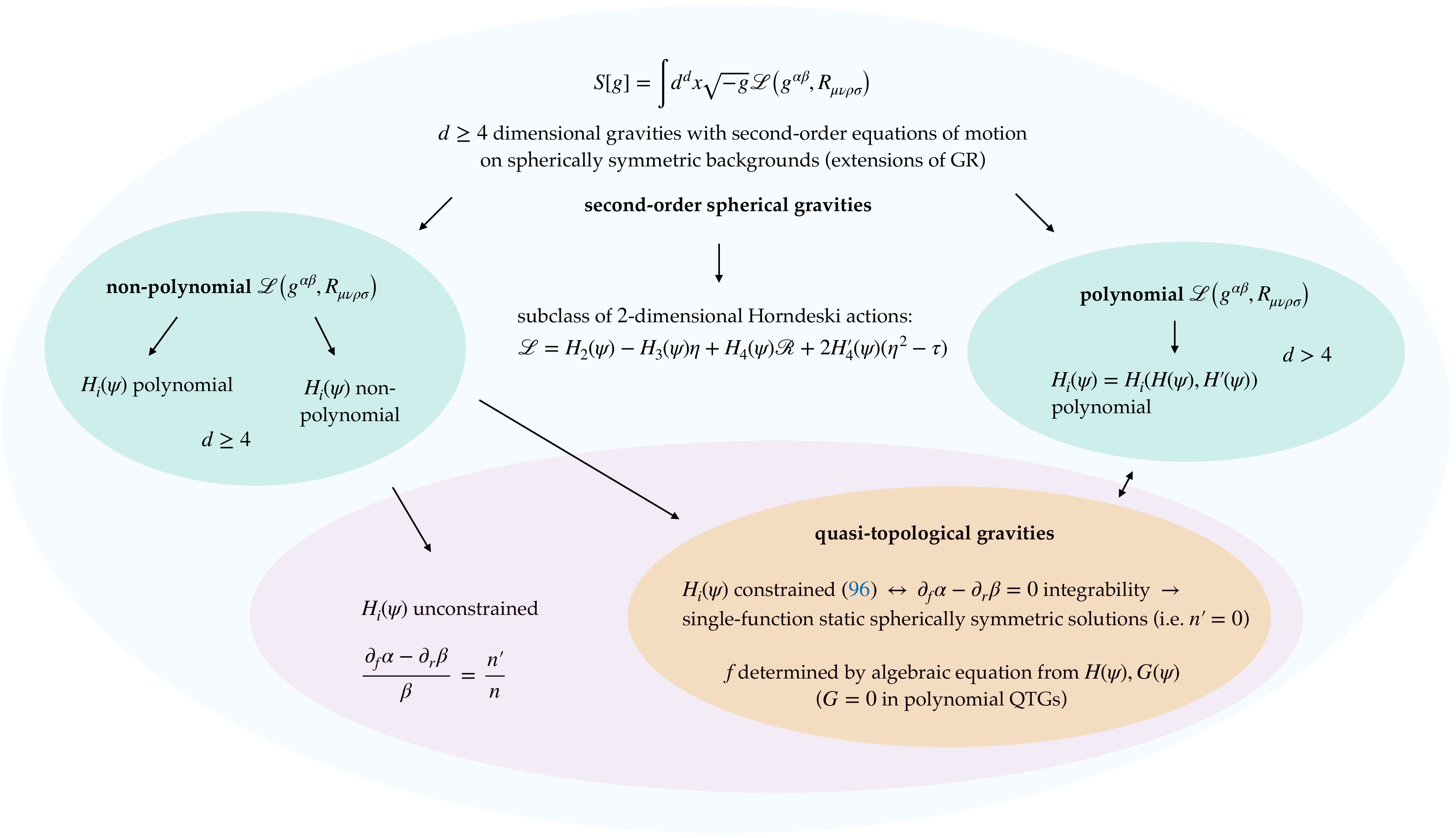}
	\caption{\label{Fig:Diagram} Landscape of $d\geq 4$ dimensional {\it second-order spherical gravities}, by which we mean here gravitational theories $\mathcal{L}\qty(g^{\alpha\beta},R_{\mu\nu\rho\sigma})$ with second-order equations of motion on spherically symmetric backgrounds. All of these theories feature reduced actions representing two-dimensional Horndeski actions of the form~\eqref{eq:SHorndeskiSubclass}. Polynomial gravities of this type only exist in $d>4$ dimensions and these are equivalent to polynomial quasi-topological gravities~\cite{Bueno:2025qjk}. In this case, the functions $H_i(\psi)$ in the reduced action are analytic functions which can be parametrized in terms of one analytic function $H(\psi)$, and the unique static spherically symmetric solutions are single-function spacetimes (i.e.~$n' = 0$) with metric function $f$ determined through an algebraic equation $H(\psi) \sim M/r^{d-1}$, where $M$ is proportional to the ADM mass of the solution~\cite{Bueno:2025qjk}. By contrast, non-polynomial gravities with second-order equations of motion on spherically symmetric backgrounds can be constructed in general dimension $d\geq 4$ by an application of the procedure described in this work. In this case, the functions $H_i(\psi)$ in the reduced action can be arbitrary, not necessarily analytic, functions. In this case, one may further distinguish between non-polynomial quasi-topological gravities satisfying the integrability constraint on the functions $H_i(\psi)$~\eqref{eq:HConstraint}, understood prior to performing any partial integrations following the reduction, which guarantees that static spherically symmetric solutions are single-function spacetimes, whereby the metric function $f$ is determined by an algebraic equation involving two functions $H(\psi)$ and $G(\psi)$ in general, cf.~\eqref{eq:EqMasterHG} in $d=4$. Generic static spherically symmetric solutions to non-polynomial second-order gravities which are not of quasi-topological type are in general spacetimes with two distinct metric functions $n$ and $f$.
	}
\end{figure}

The results and discussions of this section are summarized in Figure~\ref{Fig:Diagram}.

\section{Examples of regular black holes and corresponding gravitational actions}\label{Sec:RegularBHs}

In this section, we provide examples of different choices of characteristic function $H(\psi)$  and resulting spacetimes~\eqref{eq:MetricSSSn1} with metric function $f(r)$ generated by the algebraic equation~\eqref{eq:EqMaster}, and describe how the associated two-dimensional Horndeski action~\eqref{eq:SHorndeskiSubclass} and therefrom the four-dimensional non-polynomial quasi-topological gravitational action~\eqref{eq:SLift} can be reconstructed --- which yields this spacetime as a solution. As emphasized previously, since our construction does not rely on polynomial resummations of four-dimensional densities, each of which has a reduced action of the form~\eqref{eq:SHorndeskiSubclass} with polynomial functions $H_i(\psi)$, we able able to generate spacetimes that cannot be generated by the non-polynomial quasi-topological gravities constructed in~\cite{Bueno:2025zaj}. In particular, the function $H(\psi)$ here can be an arbitrary, not necessarily analytic, function of $\psi$.

\subsection{Hayward spacetime}\label{SecSub:Hayward}
 
The Hayward spacetime~\cite{Hayward:2005gi} is described by the line element~\eqref{eq:MetricSSSn1} where
 \ba
 f(r) &=& 1 - \frac{2Mr^2}{r^3 + 2 M l^2} \quad  \,\,\, \to \quad \,\,\, M(r,\psi) \,\,= \,\, \frac{1}{2}\frac{r^3 \psi}{1 - l^2 \psi}\,,
 \ea
 where $l$ denotes a regularization length parameter.
Therefrom we can identify the function $\Omega(r,\psi)$ and the function $H(\psi)$ in~\eqref{eq:EqMaster} as follows,
\ba\label{eq:OmegaHayward}
\Omega(r,\psi) \,\,=\,\, 4 M(r,\psi) \,\,= \,\, \frac{1}{3}r^3 H(\psi) \,\,\,\quad \text{with} \quad \,\,\, H(\psi) \,\,= \,\, \frac{6 \psi}{1 - l^2 \psi}\,,
\ea
which is in agreement with the geometric-series resummation resulting in the generating function stated in~\cite{Bueno:2024dgm}. From~\eqref{eq:OmegaHayward} we compute the functions $\alpha$ and $\beta$ according to~\eqref{eq:Omegarpsi},
\ba
\alpha(r,\psi) \,\,=\,\, \frac{2 r^2 \psi(1-3 l^2 \psi)}{\qty(1 - l^2 \psi)^2}\,\,\,\quad \text{and} \quad \,\,\, \beta(r,\psi) \,\,=\,\, - \frac{2 r}{\qty(1 - l^2 \psi)^2}\,,
\ea
and confirm that these are of the form~\eqref{eq:AlphaBetaConstrained}. One may also verify that these expressions coincide with the expressions~\eqref{eq:AlphaHaywardd}--\eqref{eq:BetaHaywardd} for $d=4$ with $\varphi = r$ and $\chi = 1 - r^2 \psi$. Thus, according to~\eqref{eq:AlphaBetaConstrained} we can identify the functions $H_2(\psi)$, $H_3(\psi)$ and $H_4(\psi)$ as follows,
\ba
H_2(\psi) &=&  \frac{2 \psi(1-3 l^2 \psi)}{\qty(1 - l^2 \psi)^2}\,,\\
H_3(\psi) &=& \frac{4 }{\qty(1 - l^2 \psi)^2}\,,\\
H_4(\psi) &=& 1 - \frac{l^2 \psi}{1 - l^2 \psi} + 2 l^2 \psi \ln\qty[\frac{1 - l^2 \psi}{\psi}]\,,
\ea
whereby we see that $H_2 $ and $H_3$ are indeed related through~\eqref{eq:H2H3}, and moreover we have used~\eqref{eq:H4Sol} with $\Lambda' \equiv 0$ to compute $H_4$. 
Therefrom we conclude that the four-dimensional gravitational action~\eqref{eq:SLiftAgain},
with functions $H_i(\mathcal{P})$ chosen as above, yields the Hayward spacetime as solution.

\subsection{Dymnikova spacetime}\label{SecSub:Dymnikova}

  The Dymnikova spacetime~\cite{Dymnikova:1992ux} is described by the line element~\eqref{eq:MetricSSS} with
 \ba
 f(r) &=& 1 - \frac{2M}{r}\qty(1 - e^{-\frac{r^3}{2 M l^2}}),
 \ea
 where $l$ denotes a regularization length parameter.  Solving for $M$ in this case involves a transcendental equation which cannot be solved with elementary functions. However, from the condition
 \begin{equation}\label{eq:omegaimpdym}
 \psi =\frac{\Omega}{2r^3}\left(1-e^{-2r^3/\Omega l^2}\right)  
 \end{equation}
one may already observe that the solution for $\Omega(r,\psi)$ will satisfy the scaling law $\Omega(\lambda r,\psi)=\lambda^3\Omega(r,\psi)$ for $\lambda\in\mathbb{R}$.
 Therefrom one may expect that the Dymnikova spacetime can be generated at least as an implicit solution to the algebraic equation~\eqref{eq:EqMaster}.\\

In fact, as pointed out in~\cite{Konoplya:2024kih}, an explicit inversion is possible in terms of the Lambert $W_0$ function satisfying the defining relation $W_0(z) = w$ when $z=w e^w$. In particular, the function $\Omega(r,\psi)$ and the function $H(\psi)$ in~\eqref{eq:EqMaster} can be expressed as follows,
\ba\label{eq:OmegaDymnikova}
\Omega(r,\psi) \,\,=\,\, 4 M(r,\psi) \,\,= \,\, \frac{1}{3}r^3 H(\psi) \,\,\,\quad \text{with} \quad \,\,\, H(\psi) \,\,= \,\, \frac{6  \psi}{1 + l^2 \psi W_0 \qty(-\frac{e^{- \frac{1}{l^2 \psi}}}{l^2 \psi})}\,.
\ea
Thus  the Dymnikova spacetime can be obtained as a solution to four-dimensional non-polynomial quasi-topological gravities.

  \subsection{Other regular black-hole spacetimes}\label{SecSub:OtherRBHs}

For completeness, in the table below we collect examples of different choices of function $H(\psi)$ and corresponding spacetimes~\eqref{eq:MetricSSSn1} with function $f(r)$ determined by the algebraic equation~\eqref{eq:EqMaster}. Higher-dimensional analogues of some of these spacetimes have been derived as solutions to quasi-topological gravities in~\cite{Bueno:2024dgm}. Our main result here is that these spacetimes can arise as vacuum solutions to purely gravitational non-polynomial quasi-topological theories in four dimensions, following the derivations and discussions presented in this and previous sections. Note that all dynamical solutions of the field equations that result from coupling the tensor defined in Eq.~\eqref{eq:GGeneralized} to a matter source~\cite{Carballo-Rubio:2025ntd} are automatically solutions of the theories discussed here when coupled to the same matter content, see~\cite{Boyanov:2025pes,Borissova:2026dlz} for discussions of such solutions.

    \begin{center}
	\begin{table}[h!]\label{Tab:Examples}
		\begin{tabular}{|c  |c | c | c|} 
			\hline
			$H(\psi)$ & $f(r)$  \\ 
			\hline\hline
			$\frac{6 \psi}{1-l^2 \psi}$ & $1 - \frac{2 M r^2}{r^3 + 2 M l^2}$ \\ 
			$\frac{6\psi}{\sqrt{1 - l^4 \psi^2}}$ & $ 1- \frac{2 M r^2}{\sqrt{r^6 + 4 l^4 M^2}} $\\
			$- \frac{6}{l^2}\ln\qty[1 -l^2 \psi]$ & $ 1 - \frac{r^2}{l^2}\qty(1 - e^{-\frac{2 M l^2}{r^3}})$ \\
			$\frac{6}{l^2} \text{arctanh} \qty(l^2 \psi)$ & $ 1 - \frac{r^2}{l^2 }\tanh\qty(\frac{2 M l^2}{r^3}) $\\
			\hline
		\end{tabular} \caption{Examples of characteristic functions $H(\psi)$ defining a particular two-dimensional Horndeski theory arising from the reduction of a four-dimensional non-polynomial gravity, and resulting metric functions $f(r)$ describing regular black-hole spacetimes for appropriately chosen values of the regularization length parameter $l$. Whenever the polynomial for $f(r)$ has different roots, we are choosing a specific one.}
	\end{table}
\end{center}
  
\section{Discussion}\label{Sec:Discussion}

We have analyzed the resolution of  the Schwarzschild singularity by a purely gravitational mechanism in four dimensions, as a result of an intrinsically non-polynomial action in curvature invariants, thereby in particular extending the central results on regular black holes from higher-dimensional polynomial quasi-topological gravities~\cite{Bueno:2024dgm}, to the physically relevant situation of four-dimensional spacetime. 

The four-dimensional theories that admit such solutions can be classified as non-polynomial quasi-topological gravities, in a doubly extended sense compared to the non-polynomial quasi-topological gravities constructed in~\cite{Bueno:2025zaj}. On the one hand, the latter are still polynomial in the primary Riemann invariant $\psi$ upon reduction on spherically symmetric backgrounds, whereas the spherically reduced actions of the non-polynomial gravities constructed here are two-dimensional Horndeski actions whose free functions $H_i$ can be arbitrary functions of $\psi$. On the other hand, we have seen that the sole constraint~\eqref{eq:HConstraint} among the functions $H_i(\psi)$ is sufficient to guarantee the existence of single-function static spherically symmetric solutions. \\

Regarding examples, we have here focused only on a particular realization of this constraint, in which these functions are parametrized in terms of a  one-variable function $H$ which determines the single metric function $f$ through an algebraic equation analogously as in polynomial quasi-topological gravities~\cite{Bueno:2025qjk} and in the non-polynomial quasi-topological gravities constructed in~\cite{Bueno:2025zaj}, where, however, $H$ does not need to be an analytic function at each finite (effective) order in the curvature. In particular, no resummation of an infinite tower of quasi-topological densities as in~\cite{Bueno:2024dgm,Bueno:2025zaj} is required to produce regular black-hole spacetimes. As a result, we have been able to explicitly construct four-dimensional gravitational actions that yield the Hayward and Dymnikova spacetimes as a solution.~\footnote{This also establishes that some of the renormalization-group improved black-hole spacetimes considered in asymptotic safety, such as the one obtained in~\cite{Bonanno:2000ep}, see, e.g., \cite{Eichhorn:2022bgu,Platania:2023srt} for reviews, can be derived from a generally covariant four-dimensional gravitational theory following the discussion of~\cite{Borissova:2026dlz}.}

For completeness, let us mention that the Hayward spacetime has been obtained as a solution to higher-dimensional quasi-topological gravities in~\cite{Bueno:2024dgm} (albeit only in odd dimensions, see, e.g., \cite{Bueno:2024zsx}), but could not be obtained in the four-dimensional non-polynomial quasi-topological gravities considered in~\cite{Bueno:2025zaj}, because the actions constructed therein use the Gauss-Bonnet invariant at order two in the curvature, and in addition combine the non-polynomial quasi-topological densities polynomially in the action, such that no contribution to the equations of motion at order two can be generated. This prevents generating the resummation series required to produce the the characteristic function that yields the Hayward spacetimes. \\

 It is interesting to notice that  an algebraic equation of the form~\eqref{eq:EqMaster} has been advocated independently from arguments based on three-dimensional spatial diffeomorphism invariance and a geometric guiding principle that ensures the absence of additional scales beyond the ADM mass regardless of relational clock fields within a canonical approach to gravity~\cite{Giesel:2025kdl}, see also~\cite{Alonso-Bardaji:2025hda} for a related work. On the other hand, it has been discussed~\cite{Frolov:2024hhe,Bueno:2024zsx,Bueno:2025tli,Frolov:2026rcm} that regular black holes arising as vacuum solutions to polynomial quasi-topological gravities, and thus satisfying the same algebraic relation, satisfy Markov's limiting curvature conjecture~\cite{Markov:1982rcm}. Our result that Eq.~\eqref{eq:EqMaster} represents a constraint satisfied in a subclass of non-polynomial theories, with Eq.~\eqref{eq:EqMasterHG} being the more general necessary and sufficient condition to guarantee that vacuum solutions are characterized by a single one-variable function, invites considerations of whether universal covariant principles responsible for characteristic structures of resulting solutions might allow to single out this subfamily of theories.\\

More generally, our discussion shows that the pure assumption that the theory possesses second-order equations of motion on static spherically symmetric backgrounds, does not single out non-polynomial quasi-topological gravities. Instead, we have shown that there are extended classes of non-polynomial gravities with second-order equations of motion on generic spherically symmetric spacetimes, but which are not of quasi-topological type, i.e., they do not satisfy the integrability condition~\eqref{eq:HConstraint} and hence they admit static spherically symmetric solutions involving two distinct metric functions. Moreover, these theories admit any of the dynamical solutions (see, e.g.,~\cite{Boyanov:2025pes,Borissova:2026dlz}) of the theories obtained by coupling two-dimensional Horndeski theory to additional matter fields that provide an effective geometrodynamic framework for $d$-dimensional gravitational fields interacting with matter, as described in~\cite{Carballo-Rubio:2025ntd}, in which a Birkhoff-type theorem holds in that vacuum solutions display a single one-parameter family of integration constants, whereby $g_{tt} g_{rr} \neq - 1$ in general.

Having addressed the possibility of deriving static spherically symmetric black holes as vacuum solutions to four-dimensional gravitational theories, a natural follow-up question is whether these can form dynamically during gravitational collapse, or more generally whether there exists a natural time-dependent generalisation of the framework and analyses presented here. This question can be addressed by coupling the associated two-dimensional reduced theory to external matter sources obtained from the spherical reduction of a four-dimensional energy-momentum tensor, and can be answered to the positive based on previous works. A solid ground to study this and related questions is provided by the second-order master field equations for spherically symmetric gravitational fields beyond general relativity~\cite{Carballo-Rubio:2025ntd}, in which the enlargement of two-dimensional Horndeski theory to include additional matter fields is interpreted as an effective geometrodynamic framework, which has been applied in particular to the formation of four-dimensional regular black holes in~\cite{Boyanov:2025pes}, 
to the analysis of the hydrostatic equilibrium of spherical stars in~\cite{Arrechea:2026ngi},  and to the description of non-singular cosmologies in~\cite{Borissova:2026klg}. Since here we have shown explicitly that there exists a four-dimensional gravitational action giving rise to the static Hayward black hole, some of the examples considered in these works, in retrospective, can be viewed as solving this particular covariant gravitational theory coupled to matter. For related discussions on the formation of regular black holes in $d\geq 5$ polynomial and $d=4$ non-polynomial quasi-topological gravities, see~\cite{Bueno:2025zaj,Bueno:2024zsx,Bueno:2025gjg}. Adopting an even broader perspective to regular black holes and alternative non-singular geometries from diverse perspectives in classical and quantum gravity, e.g.,~\cite{Biasi:2022ktq,Bonanno:2023rzk,Harada:2025cwd,Jampolski:2025lrh}, one may expect the framework of two-dimensional Horndeski theory as a reduced higher-dimensional theory to provide a rich ground for the analysis of the formation and dynamical evaolution of such geometries. See for instance~\cite{Borissova:2026dlz} for a related discussion.\\

Let us close this discussion by commenting on the viability of the non-polynomial second-order spherical gravities constructed here, in the context of an effective field theory of gravity. The theories considered here, while providing a proof of principle that regular black holes can arise as gravitational vacuum solutions, appear unnatural from the point of view of effective field theory. One aspect to this is the assumed absence of covariant derivatives in the gravitational action~\eqref{eq:S}. In fact, if the action is allowed in addition to depend on curvature invariants involving covariant derivatives, generic two-dimensional Horndeski theories relevant for the description of spherically symmetric black holes can arise from the spherical reduction of purely gravitational theories in $d\geq 4$ dimensions, thereby allowing for the interpretation of more general static spherically symmetric and asymptotically flat regular black holes as gravitational vacuum solutions~\cite{Borissova:2026krh}. However, the core aspect of non-polynomiality of the action implies that for any given two-dimensional Horndeski theory, there is an infinite set of generally covariant higher-dimensional actions which produce this theory upon reduction on spherically symmetric backgrounds, but which are otherwise inequivalent on generic backgrounds. In particular,  it is clear that such non-polynomial gravitational actions with second-order equations on spherically symmetric backgrounds may exhibit pathological properties on generic backgrounds. This is in stark contrast to polynomial pure-curvature quasi-topological gravities existing in $d\geq 5$ dimensions --- in this case, at each order $n$ in the curvature, there exist several quasi-topological densities, but all of these contribute in the same way to the equations of motion restricted to a single-function static spherically symmetric metric~\cite{Bueno:2019ycr,Bueno:2022res,Moreno:2023rfl}. Moreover, the family of polynomial curvature quasi-topological gravities in $d\geq 5$ provides a basis of polynomial densities for a pure-curvature effective field theory expansion beyond general relativity~\cite{Bueno:2019ltp}, which in particular leads to the statement that an infinite number of curvatures $n\to \infty$ in the action is required for the resolution of the Schwarzschild singularity~\cite{Bueno:2024dgm} --- a statement in favor of the generic expectation that singularity resolution is triggered by quantum effects, and that these manifest themselves through infinite towers of higher-derivative operators in the effective action. Analogous statements concerning the existence of polynomial quasi-topological densities involving covariant derivatives have, to the best of our knowledge, not been established. See, however,~\cite{Aguilar-Gutierrez:2023kfn} for a related discussion. In the case when the action and its reduction depend non-polynomially on curvature invariants, the conditions for singularity resolution and relation to effective actions motivated by effective field theory approaches to quantum gravity are not immediate, although RG-improvement in asymptotic safety grounded on the decoupling mechanism provides one such tentative connection~\cite{Borissova:2022mgd}. 
It is not excluded that some of the actions proposed here can arise as effective actions obtained by integrating out matter fields from a gravity-matter theory which admits regular black holes, although such a connection remains speculative at this stage.

\begin{acknowledgments}
J.B.~is supported by STFC Consolidated Grant ST/X000575/1 and Simons Investigator Award~690508. R.C.R.~acknowledges financial support provided by the Spanish Government through the Ram\'on y Cajal program (contract RYC2023-045894-I) and the Grant No.~PID2023-149018NB-C43 funded~by MCIN/AEI/10.13039/501100011033, and by the Junta de Andaluc\'{\i}a 
through the project FQM219 and from the Severo Ochoa grant 
CEX2021-001131-S funded by MCIN/AEI/ 10.13039/501100011033. 
\end{acknowledgments}

\appendix

\section{Zakhary-McIntosh invariants for warped-product spacetimes}\label{App:ZMInvariants}

Below we state the ZM invariants~\eqref{eq:ZMFirst}--\eqref{eq:ZMLast} evaluated for the warped-product metrics~\eqref{eq:Metric} as functions of the four combinations $\mathcal{R}$, $\eta$, $\tau$ and $\psi$ defined in~\eqref{eq:BuildingBlocks}. For this class of metrics all ZM invariants involving the dual Weyl tensor vanish identically. The remaining ones evaluate to 

\ba
\eval{\mathcal{I}_{R}}_{\eqref{eq:Metric}} &=& \mathcal{R} - 4\eta + 2 \psi\,,\label{eq:RicciScalar}\\
\eval{\mathcal{I}_{R^2}}_{\eqref{eq:Metric}} &=& \frac{1}{2}\mathcal{R}^2 - 2 \mathcal{R}\eta + 2 \eta^2 - 4 \eta\psi  + 4 \psi+2 \psi^2 \,,\\
\eval{\mathcal{I}_{C^2}}_{\eqref{eq:Metric}} &=&  \frac{1}{3}\mathcal{R}^2 + \frac{4}{3}\mathcal{R}\eta + \frac{4}{3}\mathcal{R}\psi + \frac{4}{3}\eta^2 + \frac{8}{3}\eta\psi+ \frac{4}{3} \psi^2 \,,\\
\eval{\mathcal{I}_{R^3}}_{\eqref{eq:Metric}} &=& \frac{1}{4}\mathcal{R}^3 - \frac{3}{2}\mathcal{R}^2 \eta  + 6 \mathcal{R} \tau+ 2\eta^3 +  6 \eta^2 \psi - 12\eta \tau  - 6\eta\psi^2 + 2 \psi^3 \,,\\ 
\eval{\mathcal{I}_{R^2 C}}_{\eqref{eq:Metric}} &=& -\frac{1}{12}\mathcal{R}^3 - \frac{1}{6}\mathcal{R}^2\eta + \frac{1}{6}\mathcal{R}^2 \psi -\frac{1}{3} \mathcal{R}\eta^2  + \frac{2}{3}\mathcal{R} \eta \psi + \frac{2}{3}\mathcal{R}\tau + \frac{1}{3}\mathcal{R}\psi^2 - \frac{2}{3}\eta^3 - \frac{2}{3}\eta^2 \psi  \nn\\
&+&  \frac{4}{3}\eta \tau - \frac{2}{3}\eta\psi^2 + \frac{4}{3}\tau \psi - \frac{2}{3}\psi^3 \,,\\
\eval{\mathcal{I}_{C^3}}_{\eqref{eq:Metric}} &=& \frac{1}{18}\mathcal{R}^3 + \frac{1}{3} \mathcal{R}^2 \eta + \frac{1}{3}\mathcal{R}^2 \psi + \frac{2}{3}\mathcal{R}\eta^2 + \frac{4}{3}\mathcal{R}\eta\psi + \frac{2}{3}\mathcal{R}\psi^2 + \frac{4}{9}\eta^3 + \frac{4}{3}\eta^2 \psi + \frac{4}{3}\eta\psi^2\nn\\
& +& \frac{4}{9}\psi^3 \,,\\
\eval{\mathcal{I}_{R^4}}_{\eqref{eq:Metric}} &=& \frac{1}{8}\mathcal{R}^4 - \mathcal{R}^3 \eta + 6 \mathcal{R}^2 \tau + 8 \mathcal{R} \eta^3 - 24 \mathcal{R} \eta \tau - 6 \eta^4 - 8 \eta^3 \psi + 16 \eta^2 \tau + 12 \eta^2 \psi^2\nn\\
& -& 8\eta\psi^3 + 8 \tau^2 + 2 \psi^4  \,,\\
\eval{\mathcal{I}_{R^2 C^2}}_{\eqref{eq:Metric}} &=& \frac{1}{36}\mathcal{R}^4 + \frac{1}{9} \mathcal{R}^3 \eta + \frac{1}{18} \mathcal{R}^2 \eta^2 - \frac{2}{9}\mathcal{R}^2 \eta \psi + \frac{1}{9} \mathcal{R}^2 \tau - \frac{2}{9}\mathcal{R}^2 \psi^2-\frac{2}{9}\mathcal{R}\eta^3 - \frac{2}{3}\mathcal{R}\eta^2 \psi\nn \quad \quad \quad \quad  \\
& +& \frac{4}{9}\mathcal{R}\eta \tau - \frac{4}{9}\mathcal{R} \eta\psi^2
+ \frac{4}{9}\mathcal{R}\tau\psi -\frac{2}{9}\eta^4 - \frac{4}{9}\eta^3 \psi + \frac{4}{9}\eta^2\tau 
+ \frac{2}{9}\eta^2 \psi^2 + \frac{8}{9}\eta\tau \psi
 + \frac{8}{9}\eta\psi^3 \nn\\
 &+& \frac{4}{9}\tau\psi^2 + \frac{4}{9}\psi^4\,,\quad \quad \\
\eval{\mathcal{I}_{R^4 C}}_{\eqref{eq:Metric}} &=& \frac{1}{48}\mathcal{R}^5 - \frac{1}{8}\mathcal{R}^4 \eta + \frac{1}{24}\mathcal{R}^4 \psi + \frac{1}{6}\mathcal{R}^3 \eta^2 - \frac{1}{3}\mathcal{R}^3 \tau - \frac{1}{6} \mathcal{R}^3 \psi^2 + \frac{1}{3}\mathcal{R}^2 \eta^3+ \frac{1}{3}\mathcal{R}^2 \eta^2 \psi \nn\\
&+& \frac{2}{3}\mathcal{R}^2 \eta \tau + \mathcal{R}^2 \eta\psi^2
- \frac{2}{3}\mathcal{R}^2 \tau\psi - \frac{1}{3}\mathcal{R}^2 \psi^3 + \frac{1}{3}\mathcal{R}\eta^4 - \frac{16}{3}\mathcal{R}\eta^3 \psi - \frac{4}{3}\mathcal{R}\eta^2 \tau \nn\\
&+& \frac{2}{3}\mathcal{R} \eta^2 \psi^2  + \frac{16}{3}\mathcal{R}\eta\tau \psi + \frac{4}{3}\mathcal{R}\tau^2 - \frac{4}{3}\mathcal{R}\tau\psi^2 +\frac{1}{3}\mathcal{R}\psi^4 + \frac{10}{3}\eta^5 + \frac{2}{3}\eta^4 \psi \nn\\
&-& 8 \eta^3 \tau + \frac{4}{3}\eta^3 \psi^2 - \frac{8}{3}\eta^2 \tau\psi + \frac{4}{3}\eta^2 \psi^3 + \frac{8}{3}\eta\tau^2 + \frac{8}{3}\eta\tau\psi^2 - 2 \eta \psi^4 \nn\\
&+&  \frac{8}{3}\tau^2 \psi - \frac{8}{3}\tau\psi^3 + \frac{2}{3}\psi^5\,, \\
\eval{\mathcal{I}_{R^2 C^3}}_{\eqref{eq:Metric}} &=& -\frac{1}{108}\mathcal{R}^5 - \frac{1}{18}\mathcal{R}^4 \eta - \frac{1}{54}\mathcal{R}^4 \psi - \frac{13}{108}\mathcal{R}^3 \eta^2 + \frac{1}{54}\mathcal{R}^3 \tau + \frac{2}{27} \mathcal{R}^3 \psi^2 - \frac{7}{54}\mathcal{R}^2 \eta^3 \nn\\
&+& \frac{1}{6}\mathcal{R}^2 \eta^2 \psi + \frac{1}{9}\mathcal{R}^2 \eta\tau +  \frac{4}{9}\mathcal{R}^2 \eta\psi^2 + \frac{1}{9}\mathcal{R}^2 \tau\psi + \frac{4}{27}\mathcal{R}^2 \psi^3 
- \frac{1}{9}\mathcal{R}\eta^4 \nn\\
&+& \frac{2}{27}\mathcal{R}\eta^3 \psi + \frac{2}{9}\mathcal{R}\eta^2 \tau + \frac{1}{3}\mathcal{R} \eta^2 \psi^2  + \frac{4}{9}\mathcal{R}\eta\tau \psi  + \frac{2}{9}\mathcal{R}\tau\psi^2 - \frac{4}{27}\mathcal{R}\psi^4\nn\\
& -& \frac{2}{27}\eta^5 - \frac{2}{9}\eta^4 \psi +\frac{4}{27} \eta^3 \tau - \frac{14}{27}\eta^3 \psi^2 +\frac{4}{9}\eta^2 \tau\psi - \frac{26}{27}\eta^2 \psi^3 + \frac{4}{9}\eta\tau\psi^2 \nn\\
&-& \frac{8}{9} \eta\psi^4 +  \frac{4}{27}\tau\psi^3 - \frac{8}{27}\psi^5\,.
\ea

\bibliographystyle{jhep}
\bibliography{references}

@article{Carminati:1991ddy,
	author = "Carminati, J. and McLenaghan, R. G.",
	title = "{Algebraic invariants of the Riemann tensor in a four-dimensional Lorentzian space}",
	doi = "10.1063/1.529470",
	journal = "J. Math. Phys.",
	volume = "32",
	number = "11",
	pages = "3135--3140",
	year = "1991"
}

@article{Biasi:2022ktq,
    author = "Biasi, Anxo and Evnin, Oleg and Sypsas, Spyros",
    title = "{de Sitter Bubbles from Anti{\textendash}de Sitter Fluctuations}",
    eprint = "2209.06835",
    archivePrefix = "arXiv",
    primaryClass = "gr-qc",
    doi = "10.1103/PhysRevLett.129.251104",
    journal = "Phys. Rev. Lett.",
    volume = "129",
    number = "25",
    pages = "251104",
    year = "2022"
}

@article{Bonanno:2023rzk,
    author = "Bonanno, Alfio and Malafarina, Daniele and Panassiti, Antonio",
    title = "{Dust Collapse in Asymptotic Safety: A Path to Regular Black Holes}",
    eprint = "2308.10890",
    archivePrefix = "arXiv",
    primaryClass = "gr-qc",
    doi = "10.1103/PhysRevLett.132.031401",
    journal = "Phys. Rev. Lett.",
    volume = "132",
    number = "3",
    pages = "031401",
    year = "2024"
}

@article{Harada:2025cwd,
    author = "Harada, Tomohiro and Chen, Chiang-Mei and Mandal, Rituparna",
    title = "{Singularity resolution and regular black hole formation in gravitational collapse in asymptotically safe gravity}",
    eprint = "2502.16787",
    archivePrefix = "arXiv",
    primaryClass = "gr-qc",
    reportNumber = "RUP-25-3",
    doi = "10.1103/pt9s-jqjz",
    journal = "Phys. Rev. D",
    volume = "111",
    number = "12",
    pages = "126017",
    year = "2025"
}

@article{Jampolski:2025lrh,
    author = "Jampolski, Daniel and Rezzolla, Luciano",
    title = "{On the formation of gravastars}",
    eprint = "2509.15302",
    archivePrefix = "arXiv",
    primaryClass = "gr-qc",
    month = "9",
    year = "2025"
}

@article{Frolov:2025ddw,
	author = "Frolov, Valeri P.",
	title = "{Quasitopological gravity and double-copy formalism}",
	eprint = "2512.14674",
	archivePrefix = "arXiv",
	primaryClass = "gr-qc",
	month = "12",
	year = "2025"
}

@article{Arrechea:2026ngi,
	author = "Arrechea, Julio and Carballo-Rubio, Ra{\'u}l and Visser, Matt",
	title = "{Effective geometrostatics of spherical stars beyond general relativity}",
	eprint = "2603.24269",
	archivePrefix = "arXiv",
	primaryClass = "gr-qc",
	month = "3",
	year = "2026"
}

@article{Konoplya:2024kih,
	author = "Konoplya, R. A. and Zhidenko, A.",
	title = "{Dymnikova black hole from an infinite tower of higher-curvature corrections}",
	eprint = "2404.09063",
	archivePrefix = "arXiv",
	primaryClass = "gr-qc",
	doi = "10.1016/j.physletb.2024.138945",
	journal = "Phys. Lett. B",
	volume = "856",
	pages = "138945",
	year = "2024"
}

@article{Ling:2025ncw,
	author = "Ling, Yi and Yu, Zhangping",
	title = "{Big bounce and black bounce in quasi-topological gravity}",
	eprint = "2509.00137",
	archivePrefix = "arXiv",
	primaryClass = "gr-qc",
	month = "8",
	year = "2025"
}

@article{Bonanno:2000ep,
	author = "Bonanno, Alfio and Reuter, Martin",
	title = "{Renormalization group improved black hole space-times}",
	eprint = "hep-th/0002196",
	archivePrefix = "arXiv",
	reportNumber = "INFN-CT-03-00, MZ-TH-00-04",
	doi = "10.1103/PhysRevD.62.043008",
	journal = "Phys. Rev. D",
	volume = "62",
	pages = "043008",
	year = "2000"
}

@article{Eichhorn:2022bgu,
	author = "Eichhorn, Astrid and Held, Aaron",
	title = "{Black holes in asymptotically safe gravity and beyond}",
	eprint = "2212.09495",
	archivePrefix = "arXiv",
	primaryClass = "gr-qc",
	month = "12",
	year = "2022"
}

@article{Borissova:2022mgd,
	author = "Borissova, Johanna N. and Platania, Alessia",
	title = "{Formation and evaporation of quantum black holes from the decoupling mechanism in quantum gravity}",
	eprint = "2210.01138",
	archivePrefix = "arXiv",
	primaryClass = "gr-qc",
	reportNumber = "NORDITA 2022-069",
	doi = "10.1007/JHEP03(2023)046",
	journal = "JHEP",
	volume = "03",
	pages = "046",
	year = "2023"
}

@article{Borissova:2026dlz,
    author = "Borissova, Johanna and Carballo-Rubio, Ra{\'u}l",
    title = "{Effective geometrodynamics for renormalization-group improved black-hole spacetimes in spherical symmetry}",
    eprint = "2601.17115",
    archivePrefix = "arXiv",
    primaryClass = "gr-qc",
    reportNumber = "Imperial/TP/2026/JB/01",
    doi = "10.1088/1475-7516/2026/05/023",
    journal = "JCAP",
    volume = "05",
    pages = "023",
    year = "2026"
}

@article{Zakhary:1997xas,
	author = "Zakhary, E. and Mcintosh, C. B. G.",
	title = "{A Complete Set of Riemann Invariants}",
	doi = "10.1023/a:1018851201784",
	journal = "Gen. Rel. Grav.",
	volume = "29",
	number = "5",
	pages = "539--581",
	year = "1997"
}

@article{Held:2021vwd,
	author = "Held, Aaron",
	title = "{Invariant Renormalization-Group improvement}",
	eprint = "2105.11458",
	archivePrefix = "arXiv",
	primaryClass = "gr-qc",
	reportNumber = "Imperial/TP/2021/AH/04",
	month = "5",
	year = "2021"
}

@article{Borissova:2023kzq,
	author = "Borissova, Johanna N.",
	title = "{Suppression of spacetime singularities in quantum gravity}",
	eprint = "2309.05695",
	archivePrefix = "arXiv",
	primaryClass = "gr-qc",
	doi = "10.1088/1361-6382/ad46c0",
	journal = "Class. Quant. Grav.",
	volume = "41",
	number = "12",
	pages = "127002",
	year = "2024"
}

@article{Overduin:2020aiq,
	author = "Overduin, James and Coplan, Max and Wilcomb, Kielan and Henry, Richard Conn",
	title = "{Curvature Invariants for Charged and Rotating Black Holes}",
	doi = "10.3390/universe6020022",
	journal = "Universe",
	volume = "6",
	number = "2",
	pages = "22",
	year = "2020"
}

@article{Palais:1979rca,
	author = "Palais, Richard S.",
	title = "{The principle of symmetric criticality}",
	doi = "10.1007/BF01941322",
	journal = "Commun. Math. Phys.",
	volume = "69",
	number = "1",
	pages = "19--30",
	year = "1979"
}

@article{Deser:2003up,
	author = "Deser, Stanley and Tekin, Bayram",
	title = "{Shortcuts to high symmetry solutions in gravitational theories}",
	eprint = "gr-qc/0306114",
	archivePrefix = "arXiv",
	reportNumber = "BRX-TH-520",
	doi = "10.1088/0264-9381/20/22/011",
	journal = "Class. Quant. Grav.",
	volume = "20",
	pages = "4877--4884",
	year = "2003"
}

@article{Bueno:2019ycr,
	author = "Bueno, Pablo and Cano, Pablo A. and Hennigar, Robie A.",
	title = "{(Generalized) quasi-topological gravities at all orders}",
	eprint = "1909.07983",
	archivePrefix = "arXiv",
	primaryClass = "hep-th",
	reportNumber = "IFT-UAM/CSIC-19-124",
	doi = "10.1088/1361-6382/ab5410",
	journal = "Class. Quant. Grav.",
	volume = "37",
	number = "1",
	pages = "015002",
	year = "2020"
}

@article{Bueno:2019ltp,
	author = "Bueno, Pablo and Cano, Pablo A. and Moreno, Javier and Murcia, {\'A}ngel",
	title = "{All higher-curvature gravities as Generalized quasi-topological gravities}",
	eprint = "1906.00987",
	archivePrefix = "arXiv",
	primaryClass = "hep-th",
	doi = "10.1007/JHEP11(2019)062",
	journal = "JHEP",
	volume = "11",
	pages = "062",
	year = "2019"
}

@article{Bueno:2022res,
	author = "Bueno, Pablo and Cano, Pablo A. and Hennigar, Robie A. and Lu, Mengqi and Moreno, Javier",
	title = "{Generalized quasi-topological gravities: the whole shebang}",
	eprint = "2203.05589",
	archivePrefix = "arXiv",
	primaryClass = "hep-th",
	reportNumber = "CERN-TH-2022-038",
	doi = "10.1088/1361-6382/aca236",
	journal = "Class. Quant. Grav.",
	volume = "40",
	number = "1",
	pages = "015004",
	year = "2023"
}

@article{Bueno:2024dgm,
	author = "Bueno, Pablo and Cano, Pablo A. and Hennigar, Robie A.",
	title = "{Regular black holes from pure gravity}",
	eprint = "2403.04827",
	archivePrefix = "arXiv",
	primaryClass = "gr-qc",
	doi = "10.1016/j.physletb.2025.139260",
	journal = "Phys. Lett. B",
	volume = "861",
	pages = "139260",
	year = "2025"
}

@article{Bueno:2025zaj,
    author = "Bueno, Pablo and Cano, Pablo A. and Hennigar, Robie A. and Murcia, {\'A}ngel J.",
    title = "{Regular black hole formation in four-dimensional nonpolynomial gravities}",
    eprint = "2509.19016",
    archivePrefix = "arXiv",
    primaryClass = "gr-qc",
    doi = "10.1103/8f3j-zcxh",
    journal = "Phys. Rev. D",
    volume = "113",
    number = "2",
    pages = "024019",
    year = "2026"
}

@article{Bueno:2024zsx,
	author = "Bueno, Pablo and Cano, Pablo A. and Hennigar, Robie A. and Murcia, {\'A}ngel J.",
	title = "{Regular black holes from thin-shell collapse}",
	eprint = "2412.02740",
	archivePrefix = "arXiv",
	primaryClass = "gr-qc",
	doi = "10.1103/PhysRevD.111.104009",
	journal = "Phys. Rev. D",
	volume = "111",
	number = "10",
	pages = "104009",
	year = "2025"
}

@article{Moreno:2023rfl,
	author = "Moreno, Javier and Murcia, {\'A}ngel J.",
	title = "{Classification of generalized quasitopological gravities}",
	eprint = "2304.08510",
	archivePrefix = "arXiv",
	primaryClass = "gr-qc",
	doi = "10.1103/PhysRevD.108.044016",
	journal = "Phys. Rev. D",
	volume = "108",
	number = "4",
	pages = "044016",
	year = "2023"
}

@article{Bueno:2025qjk,
    author = "Bueno, Pablo and Hennigar, Robie A. and Murcia, {\'A}ngel J.",
    title = "{Birkhoff implies quasi-topological}",
    eprint = "2510.25823",
    archivePrefix = "arXiv",
    primaryClass = "gr-qc",
    doi = "10.1088/1361-6382/ae6600",
    journal = "Class. Quant. Grav.",
    volume = "43",
    number = "9",
    pages = "095020",
    year = "2026"
}

@article{Boyanov:2025pes,
    author = "Boyanov, Valentin and Carballo-Rubio, Ra{\'u}l",
    title = "{Regular Vaidya Solutions of Effective Gravitational Theories}",
    eprint = "2506.14875",
    archivePrefix = "arXiv",
    primaryClass = "gr-qc",
    doi = "10.1103/3ng1-y6vs",
    journal = "Phys. Rev. Lett.",
    volume = "136",
    number = "17",
    pages = "171403",
    year = "2026"
}

@article{Carballo-Rubio:2025ntd,
	author = "Carballo-Rubio, Ra{\'u}l",
	title = "{Master field equations for spherically symmetric gravitational fields beyond general relativity}",
	eprint = "2507.15920",
	archivePrefix = "arXiv",
	primaryClass = "gr-qc",
	doi = "10.1038/s41467-026-69035-6",
	journal = "Nature Commun.",
	volume = "17",
	number = "1",
	pages = "1399",
	year = "2026"
}

@article{Borissova:2026krh,
	author = "Borissova, Johanna",
	title = "{All $2D$ generalised dilaton theories from $d\geq 4$ gravities}",
	eprint = "2603.06786",
	archivePrefix = "arXiv",
	primaryClass = "hep-th",
	reportNumber = "Imperial/TP/2026/JB/03",
	month = "3",
	year = "2026"
}

@article{Aguilar-Gutierrez:2023kfn,
	author = "Aguilar-Gutierrez, Sergio E. and Bueno, Pablo and Cano, Pablo A. and Hennigar, Robie A. and Llorens, Quim",
	title = "{Aspects of higher-curvature gravities with covariant derivatives}",
	eprint = "2310.09333",
	archivePrefix = "arXiv",
	primaryClass = "hep-th",
	doi = "10.1103/PhysRevD.108.124075",
	journal = "Phys. Rev. D",
	volume = "108",
	number = "12",
	pages = "124075",
	year = "2023"
}

@article{Bueno:2025gjg,
	author = "Bueno, Pablo and Cano, Pablo A. and Hennigar, Robie A. and Murcia, {\'A}ngel J. and Vicente-Cano, Aitor",
	title = "{Regular black holes from Oppenheimer-Snyder collapse}",
	eprint = "2505.09680",
	archivePrefix = "arXiv",
	primaryClass = "gr-qc",
	doi = "10.1103/qrbb-mdvm",
	journal = "Phys. Rev. D",
	volume = "112",
	number = "6",
	pages = "064039",
	year = "2025"
}

@article{Borissova:2026klg,
	author = "Borissova, Johanna and Magueijo, Jo{\~a}o",
	title = "{Modified Friedmann equations and non-singular cosmologies in $d=4$ non-polynomial quasi-topological gravities}",
	eprint = "2603.17654",
	archivePrefix = "arXiv",
	primaryClass = "gr-qc",
	reportNumber = "Imperial/TP/2026/JB/04",
	month = "3",
	year = "2026"
}

@article{Horndeski:1974wa,
	author = "Horndeski, Gregory Walter",
	title = "{Second-order scalar-tensor field equations in a four-dimensional space}",
	doi = "10.1007/BF01807638",
	journal = "Int. J. Theor. Phys.",
	volume = "10",
	pages = "363--384",
	year = "1974"
}

@article{Kobayashi:2019hrl,
	author = "Kobayashi, Tsutomu",
	title = "{Horndeski theory and beyond: a review}",
	eprint = "1901.07183",
	archivePrefix = "arXiv",
	primaryClass = "gr-qc",
	reportNumber = "RUP-19-3",
	doi = "10.1088/1361-6633/ab2429",
	journal = "Rept. Prog. Phys.",
	volume = "82",
	number = "8",
	pages = "086901",
	year = "2019"
}

@article{Fels:2001rv,
	author = "Fels, Mark E. and Torre, Charles G.",
	title = "{The Principle of symmetric criticality in general relativity}",
	eprint = "gr-qc/0108033",
	archivePrefix = "arXiv",
	doi = "10.1088/0264-9381/19/4/303",
	journal = "Class. Quant. Grav.",
	volume = "19",
	pages = "641--676",
	year = "2002"
}

@article{Oliva:2010eb,
	author = "Oliva, Julio and Ray, Sourya",
	title = "{A new cubic theory of gravity in five dimensions: Black hole, Birkhoff's theorem and C-function}",
	eprint = "1003.4773",
	archivePrefix = "arXiv",
	primaryClass = "gr-qc",
	reportNumber = "CECS-PHY-10-03",
	doi = "10.1088/0264-9381/27/22/225002",
	journal = "Class. Quant. Grav.",
	volume = "27",
	pages = "225002",
	year = "2010"
}

@article{Myers:2010ru,
	author = "Myers, Robert C. and Robinson, Brandon",
	title = "{Black Holes in Quasi-topological Gravity}",
	eprint = "1003.5357",
	archivePrefix = "arXiv",
	primaryClass = "gr-qc",
	doi = "10.1007/JHEP08(2010)067",
	journal = "JHEP",
	volume = "08",
	pages = "067",
	year = "2010"
}

@article{Dehghani:2011vu,
	author = "Dehghani, M. H. and Bazrafshan, A. and Mann, R. B. and Mehdizadeh, M. R. and Ghanaatian, M. and Vahidinia, M. H.",
	title = "{Black Holes in Quartic Quasitopological Gravity}",
	eprint = "1109.4708",
	archivePrefix = "arXiv",
	primaryClass = "hep-th",
	doi = "10.1103/PhysRevD.85.104009",
	journal = "Phys. Rev. D",
	volume = "85",
	pages = "104009",
	year = "2012"
}

@article{Cisterna:2017umf,
	author = "Cisterna, Adolfo and Guajardo, Luis and Hassaine, Mokhtar and Oliva, Julio",
	title = "{Quintic quasi-topological gravity}",
	eprint = "1702.04676",
	archivePrefix = "arXiv",
	primaryClass = "hep-th",
	doi = "10.1007/JHEP04(2017)066",
	journal = "JHEP",
	volume = "04",
	pages = "066",
	year = "2017"
}

@article{Colleaux:2017ibe,
	author = "Coll{\'e}aux, Aimeric and Chinaglia, Stefano and Zerbini, Sergio",
	title = "{Nonpolynomial Lagrangian approach to regular black holes}",
	eprint = "1712.03730",
	archivePrefix = "arXiv",
	primaryClass = "gr-qc",
	doi = "10.1142/S0218271818300021",
	journal = "Int. J. Mod. Phys. D",
	volume = "27",
	number = "03",
	pages = "1830002",
	year = "2018"
}

@article{Deser:2005pc,
	author = "Deser, Stanley and Ryzhov, A. V.",
	title = "{Curvature invariants of static spherically symmetric geometries}",
	eprint = "gr-qc/0505039",
	archivePrefix = "arXiv",
	reportNumber = "BRX-TH-563",
	doi = "10.1088/0264-9381/22/16/012",
	journal = "Class. Quant. Grav.",
	volume = "22",
	pages = "3315--3324",
	year = "2005"
}

@article{Taves:2014laa,
    author = "Taves, Tim and Kunstatter, Gabor",
    title = "{Modelling the evaporation of nonsingular black holes}",
    eprint = "1408.1444",
    archivePrefix = "arXiv",
    primaryClass = "gr-qc",
    doi = "10.1103/PhysRevD.90.124062",
    journal = "Phys. Rev. D",
    volume = "90",
    number = "12",
    pages = "124062",
    year = "2014"
}

@article{Kunstatter:2015vxa,
	author = "Kunstatter, Gabor and Maeda, Hideki and Taves, Tim",
	title = "{New 2D dilaton gravity for nonsingular black holes}",
	eprint = "1509.06746",
	archivePrefix = "arXiv",
	primaryClass = "gr-qc",
	doi = "10.1088/0264-9381/33/10/105005",
	journal = "Class. Quant. Grav.",
	volume = "33",
	number = "10",
	pages = "105005",
	year = "2016"
}

@article{Takahashi:2018yzc,
	author = "Takahashi, Kazufumi and Kobayashi, Tsutomu",
	title = "{Generalized 2D dilaton gravity and kinetic gravity braiding}",
	eprint = "1812.08847",
	archivePrefix = "arXiv",
	primaryClass = "gr-qc",
	reportNumber = "RUP-18-35",
	doi = "10.1088/1361-6382/ab1355",
	journal = "Class. Quant. Grav.",
	volume = "36",
	number = "9",
	pages = "095003",
	year = "2019"
}

@article{Deffayet:2011gz,
	author = "Deffayet, C. and Gao, Xian and Steer, D. A. and Zahariade, G.",
	title = "{From k-essence to generalised Galileons}",
	eprint = "1103.3260",
	archivePrefix = "arXiv",
	primaryClass = "hep-th",
	doi = "10.1103/PhysRevD.84.064039",
	journal = "Phys. Rev. D",
	volume = "84",
	pages = "064039",
	year = "2011"
}

@article{Deffayet:2010qz,
	author = "Deffayet, Cedric and Pujolas, Oriol and Sawicki, Ignacy and Vikman, Alexander",
	title = "{Imperfect Dark Energy from Kinetic Gravity Braiding}",
	eprint = "1008.0048",
	archivePrefix = "arXiv",
	primaryClass = "hep-th",
	reportNumber = "CERN-PH-TH-2010-166",
	doi = "10.1088/1475-7516/2010/10/026",
	journal = "JCAP",
	volume = "10",
	pages = "026",
	year = "2010"
}

@article{Fernandes:2022zrq,
	author = "Fernandes, Pedro G. S. and Carrilho, Pedro and Clifton, Timothy and Mulryne, David J.",
	title = "{The 4D Einstein{\textendash}Gauss{\textendash}Bonnet theory of gravity: a review}",
	eprint = "2202.13908",
	archivePrefix = "arXiv",
	primaryClass = "gr-qc",
	doi = "10.1088/1361-6382/ac500a",
	journal = "Class. Quant. Grav.",
	volume = "39",
	number = "6",
	pages = "063001",
	year = "2022"
}

@article{Glavan:2019inb,
	author = "Glavan, Dra{\v{z}}en and Lin, Chunshan",
	title = "{Einstein-Gauss-Bonnet Gravity in Four-Dimensional Spacetime}",
	eprint = "1905.03601",
	archivePrefix = "arXiv",
	primaryClass = "gr-qc",
	reportNumber = "CP3-19-24",
	doi = "10.1103/PhysRevLett.124.081301",
	journal = "Phys. Rev. Lett.",
	volume = "124",
	number = "8",
	pages = "081301",
	year = "2020"
}

@article{Lovelock:1970zsf,
	author = "Lovelock, David",
	title = "{Divergence-free tensorial concomitants}",
	doi = "10.1007/BF01817753",
	journal = "Aequat. Math.",
	volume = "4",
	number = "1",
	pages = "127--138",
	year = "1970"
}

@article{Lovelock:1971yv,
	author = "Lovelock, D.",
	title = "{The Einstein tensor and its generalizations}",
	doi = "10.1063/1.1665613",
	journal = "J. Math. Phys.",
	volume = "12",
	pages = "498--501",
	year = "1971"
}

@article{Lanczos:1938sf,
	author = "Lanczos, Cornelius",
	title = "{A Remarkable property of the Riemann-Christoffel tensor in four dimensions}",
	doi = "10.2307/1968467",
	journal = "Annals Math.",
	volume = "39",
	pages = "842--850",
	year = "1938"
}

@article{Hayward:2005gi,
	author = "Hayward, Sean A.",
	title = "{Formation and evaporation of regular black holes}",
	eprint = "gr-qc/0506126",
	archivePrefix = "arXiv",
	doi = "10.1103/PhysRevLett.96.031103",
	journal = "Phys. Rev. Lett.",
	volume = "96",
	pages = "031103",
	year = "2006"
}

@article{Penrose:1964wq,
	author = "Penrose, Roger",
	title = "{Gravitational collapse and space-time singularities}",
	doi = "10.1103/PhysRevLett.14.57",
	journal = "Phys. Rev. Lett.",
	volume = "14",
	pages = "57--59",
	year = "1965"
}

@article{Alonso-Bardaji:2025hda,
    author = "Alonso-Bardaji, Asier and Brizuela, David",
    title = "{Dynamical theory for spherical black holes in modified gravity}",
    eprint = "2507.19380",
    archivePrefix = "arXiv",
    primaryClass = "gr-qc",
    doi = "10.1103/ttfy-yjdh",
    journal = "Phys. Rev. D",
    volume = "112",
    number = "10",
    pages = "104036",
    year = "2025"
}

@article{Bardeen:1968bh,
	author = "Bardeen, J..",
	title = "{Non-singular general-relativistic gravitational collapse}",
	journal = "Abstracts of GR5 --- the 5th international conference on gravitation and the theory of relativity, 
	eds. V.~A.~Fock et al. (Tbilisi University Press, Tbilisi, Georgia,  former USSR)",
	year = "1968",
	pages = "174--175"
}

@book{Hawking:1973uf,
	author = "Hawking, Stephen W. and Ellis, George F. R.",
	title = "{The Large Scale Structure of Space-Time}",
	doi = "10.1017/9781009253161",
	isbn = "978-1-009-25316-1, 978-1-009-25315-4, 978-0-521-20016-5, 978-0-521-09906-6, 978-0-511-82630-6, 978-0-521-09906-6",
	publisher = "Cambridge University Press",
	series = "Cambridge Monographs on Mathematical Physics",
	month = "2",
	year = "2023"
}

@article{Dymnikova:1992ux,
	author = "Dymnikova, I.",
	title = "{Vacuum nonsingular black hole}",
	doi = "10.1007/BF00760226",
	journal = "Gen. Rel. Grav.",
	volume = "24",
	pages = "235--242",
	year = "1992"
}

@article{Borissova:2025hmj,
    author = "Borissova, Johanna and Liberati, Stefano and Visser, Matt",
    title = "{Timelike convergence condition in regular black-hole spacetimes with (anti{\textendash})de Sitter core}",
    eprint = "2509.08590",
    archivePrefix = "arXiv",
    primaryClass = "gr-qc",
    doi = "10.1103/rrc9-g1sv",
    journal = "Phys. Rev. D",
    volume = "112",
    number = "10",
    pages = "104072",
    year = "2025"
}

@article{Ayon-Beato:2000mjt,
	author = "Ayon-Beato, Eloy and Garcia, Alberto",
	title = "{The Bardeen model as a nonlinear magnetic monopole}",
	eprint = "gr-qc/0009077",
	archivePrefix = "arXiv",
	doi = "10.1016/S0370-2693(00)01125-4",
	journal = "Phys. Lett. B",
	volume = "493",
	pages = "149--152",
	year = "2000"
}

@article{Fernandes:2020rpa,
	author = "Fernandes, Pedro G. S.",
	title = "{Charged black holes in AdS spaces in 4D Einstein Gauss-Bonnet gravity}",
	eprint = "2003.05491",
	archivePrefix = "arXiv",
	primaryClass = "gr-qc",
	doi = "10.1016/j.physletb.2020.135468",
	journal = "Phys. Lett. B",
	volume = "805",
	pages = "135468",
	year = "2020"
}

@article{Konoplya:2020qqh,
	author = "Konoplya, R. A. and Zhidenko, A.",
	title = "{Black holes in the four-dimensional Einstein-Lovelock gravity}",
	eprint = "2003.07788",
	archivePrefix = "arXiv",
	primaryClass = "gr-qc",
	doi = "10.1103/PhysRevD.101.084038",
	journal = "Phys. Rev. D",
	volume = "101",
	number = "8",
	pages = "084038",
	year = "2020"
}

@article{Ghosh:2020vpc,
	author = "Ghosh, Sushant G. and Maharaj, Sunil D.",
	title = "{Radiating black holes in the novel 4D Einstein{\textendash}Gauss{\textendash}Bonnet gravity}",
	eprint = "2003.09841",
	archivePrefix = "arXiv",
	primaryClass = "gr-qc",
	doi = "10.1016/j.dark.2020.100687",
	journal = "Phys. Dark Univ.",
	volume = "30",
	pages = "100687",
	year = "2020"
}

@article{Ghosh:2020syx,
	author = "Ghosh, Sushant G. and Kumar, Rahul",
	title = "{Generating black holes in $4D$ Einstein-Gauss-Bonnet gravity}",
	eprint = "2003.12291",
	archivePrefix = "arXiv",
	primaryClass = "gr-qc",
	doi = "10.1088/1361-6382/abc134",
	journal = "Class. Quant. Grav.",
	volume = "37",
	number = "24",
	pages = "245008",
	year = "2020"
}

@article{Konoplya:2020juj,
	author = "Konoplya, R. A. and Zhidenko, A.",
	title = "{(In)stability of black holes in the $4D$ Einstein{\textendash}Gauss{\textendash}Bonnet and Einstein{\textendash}Lovelock gravities}",
	eprint = "2003.12492",
	archivePrefix = "arXiv",
	primaryClass = "gr-qc",
	doi = "10.1016/j.dark.2020.100697",
	journal = "Phys. Dark Univ.",
	volume = "30",
	pages = "100697",
	year = "2020"
}

@article{Kumar:2020owy,
	author = "Kumar, Rahul and Ghosh, Sushant G.",
	title = "{Rotating black holes in $4D$ Einstein-Gauss-Bonnet gravity and its shadow}",
	eprint = "2003.08927",
	archivePrefix = "arXiv",
	primaryClass = "gr-qc",
	doi = "10.1088/1475-7516/2020/07/053",
	journal = "JCAP",
	volume = "07",
	pages = "053",
	year = "2020"
}

@article{Kumar:2020uyz,
	author = "Kumar, Arun and Walia, Rahul Kumar and Ghosh, Sushant G.",
	title = "{Bardeen Black Holes in the Regularized 4D Einstein{\textendash}Gauss{\textendash}Bonnet Gravity}",
	eprint = "2003.13104",
	archivePrefix = "arXiv",
	primaryClass = "gr-qc",
	doi = "10.3390/universe8040232",
	journal = "Universe",
	volume = "8",
	number = "4",
	pages = "232",
	year = "2022"
}

@article{Kumar:2020bqf,
	author = "Kumar, Arun and Singh, Dharm Veer and Ghosh, Sushant G.",
	title = "{Hayward black holes in Einstein{\textendash}Gauss{\textendash}Bonnet gravity}",
	eprint = "2003.14016",
	archivePrefix = "arXiv",
	primaryClass = "gr-qc",
	doi = "10.1016/j.aop.2020.168214",
	journal = "Annals Phys.",
	volume = "419",
	pages = "168214",
	year = "2020"
}

@article{Ziprick:2010vb,
    author = "Ziprick, Jonathan and Kunstatter, Gabor",
    title = "{Quantum Corrected Spherical Collapse: A Phenomenological Framework}",
    eprint = "1004.0525",
    archivePrefix = "arXiv",
    primaryClass = "gr-qc",
    doi = "10.1103/PhysRevD.82.044031",
    journal = "Phys. Rev. D",
    volume = "82",
    pages = "044031",
    year = "2010"
}

@article{Konoplya:2020cbv,
	author = "Konoplya, Roman A. and Zinhailo, Antonina F.",
	title = "{Grey-body factors and Hawking radiation of black holes in $4D$ Einstein-Gauss-Bonnet gravity}",
	eprint = "2004.02248",
	archivePrefix = "arXiv",
	primaryClass = "gr-qc",
	doi = "10.1016/j.physletb.2020.135793",
	journal = "Phys. Lett. B",
	volume = "810",
	pages = "135793",
	year = "2020"
}

@article{Malafarina:2020pvl,
	author = "Malafarina, Daniele and Toshmatov, Bobir and Dadhich, Naresh",
	title = "{Dust collapse in 4D Einstein{\textendash}Gauss{\textendash}Bonnet gravity}",
	eprint = "2004.07089",
	archivePrefix = "arXiv",
	primaryClass = "gr-qc",
	doi = "10.1016/j.dark.2020.100598",
	journal = "Phys. Dark Univ.",
	volume = "30",
	pages = "100598",
	year = "2020"
}

@article{Yang:2020jno,
	author = "Yang, Ke and Gu, Bao-Min and Wei, Shao-Wen and Liu, Yu-Xiao",
	title = "{Born{\textendash}Infeld black holes in 4D Einstein{\textendash}Gauss{\textendash}Bonnet gravity}",
	eprint = "2004.14468",
	archivePrefix = "arXiv",
	primaryClass = "gr-qc",
	doi = "10.1140/epjc/s10052-020-8246-6",
	journal = "Eur. Phys. J. C",
	volume = "80",
	number = "7",
	pages = "662",
	year = "2020"
}

@article{Feng:2020duo,
	author = "Feng, Jia-Xi and Gu, Bao-Min and Shu, Fu-Wen",
	title = "{Theoretical and observational constraints on regularized 4$D$ Einstein-Gauss-Bonnet gravity}",
	eprint = "2006.16751",
	archivePrefix = "arXiv",
	primaryClass = "gr-qc",
	doi = "10.1103/PhysRevD.103.064002",
	journal = "Phys. Rev. D",
	volume = "103",
	pages = "064002",
	year = "2021"
}

@article{Gurses:2020ofy,
	author = {G{\"u}rses, Metin and {\c{S}}i{\c{s}}man, Tahsin {\c{C}}a{\u{g}}r{\i} and Tekin, Bayram},
	title = "{Is there a novel Einstein{\textendash}Gauss{\textendash}Bonnet theory in four dimensions?}",
	eprint = "2004.03390",
	archivePrefix = "arXiv",
	primaryClass = "gr-qc",
	doi = "10.1140/epjc/s10052-020-8200-7",
	journal = "Eur. Phys. J. C",
	volume = "80",
	number = "7",
	pages = "647",
	year = "2020"
}

@article{Gurses:2020rxb,
	author = "Gurses, Metin and {\c{S}}i{\c{s}}man, Tahsin {\c{C}}a{\u{g}}r{\i} and Tekin, Bayram",
	title = "{Comment on ''Einstein-Gauss-Bonnet Gravity in 4-Dimensional Space-Time''}",
	eprint = "2009.13508",
	archivePrefix = "arXiv",
	primaryClass = "gr-qc",
	doi = "10.1103/PhysRevLett.125.149001",
	journal = "Phys. Rev. Lett.",
	volume = "125",
	number = "14",
	pages = "149001",
	year = "2020"
}

@article{Arrechea:2020evj,
	author = "Arrechea, Julio and Delhom, Adri{\`a} and Jim{\'e}nez-Cano, Alejandro",
	title = "{Inconsistencies in four-dimensional Einstein-Gauss-Bonnet gravity}",
	eprint = "2004.12998",
	archivePrefix = "arXiv",
	primaryClass = "gr-qc",
	doi = "10.1088/1674-1137/abc1d4",
	journal = "Chin. Phys. C",
	volume = "45",
	number = "1",
	pages = "013107",
	year = "2021"
}

@article{Arrechea:2020gjw,
	author = "Arrechea, Julio and Delhom, Adri{\`a} and Jim{\'e}nez-Cano, Alejandro",
	title = "{Comment on {\textquotedblleft}Einstein-Gauss-Bonnet Gravity in Four-Dimensional Spacetime{\textquotedblright}}",
	eprint = "2009.10715",
	archivePrefix = "arXiv",
	primaryClass = "gr-qc",
	doi = "10.1103/PhysRevLett.125.149002",
	journal = "Phys. Rev. Lett.",
	volume = "125",
	number = "14",
	pages = "149002",
	year = "2020"
}

@article{Bonifacio:2020vbk,
	author = "Bonifacio, James and Hinterbichler, Kurt and Johnson, Laura A.",
	title = "{Amplitudes and 4D Gauss-Bonnet Theory}",
	eprint = "2004.10716",
	archivePrefix = "arXiv",
	primaryClass = "hep-th",
	doi = "10.1103/PhysRevD.102.024029",
	journal = "Phys. Rev. D",
	volume = "102",
	number = "2",
	pages = "024029",
	year = "2020"
}

@article{Ai:2020peo,
	author = "Ai, Wen-Yuan",
	title = "{A note on the novel 4D Einstein{\textendash}Gauss{\textendash}Bonnet gravity}",
	eprint = "2004.02858",
	archivePrefix = "arXiv",
	primaryClass = "gr-qc",
	reportNumber = "CP3-20-16",
	doi = "10.1088/1572-9494/aba242",
	journal = "Commun. Theor. Phys.",
	volume = "72",
	number = "9",
	pages = "095402",
	year = "2020"
}

@article{Mahapatra:2020rds,
	author = "Mahapatra, Subhash",
	title = "{A note on the total action of 4D Gauss{\textendash}Bonnet theory}",
	eprint = "2004.09214",
	archivePrefix = "arXiv",
	primaryClass = "gr-qc",
	doi = "10.1140/epjc/s10052-020-08568-6",
	journal = "Eur. Phys. J. C",
	volume = "80",
	number = "10",
	pages = "992",
	year = "2020"
}

@article{Hohmann:2020cor,
	author = "Hohmann, Manuel and Pfeifer, Christian and Voicu, Nicoleta",
	title = "{Canonical variational completion and 4D Gauss-Bonnet gravity}",
	eprint = "2009.05459",
	archivePrefix = "arXiv",
	primaryClass = "gr-qc",
	doi = "10.1140/epjp/s13360-021-01153-0",
	journal = "Eur. Phys. J. Plus",
	volume = "136",
	number = "2",
	pages = "180",
	year = "2021"
}

@article{Cao:2021nng,
	author = "Cao, Li-Ming and Wu, Liang-Bi",
	title = "{On the {\textquotedblleft}Einstein{\textendash}Gauss{\textendash}Bonnet gravity in four dimension{\textquotedblright}}",
	eprint = "2103.09612",
	archivePrefix = "arXiv",
	primaryClass = "gr-qc",
	reportNumber = "ICTS-USTC/PCFT-21-11",
	doi = "10.1140/epjc/s10052-022-10079-5",
	journal = "Eur. Phys. J. C",
	volume = "82",
	number = "2",
	pages = "124",
	year = "2022"
}

@article{Deser:2007za,
	author = "Deser, Stanley and Sarioglu, Ozgur and Tekin, Bayram",
	title = "{Spherically symmetric solutions of Einstein + non-polynomial gravities}",
	eprint = "0705.1669",
	archivePrefix = "arXiv",
	primaryClass = "gr-qc",
	doi = "10.1007/s10714-007-0508-1",
	journal = "Gen. Rel. Grav.",
	volume = "40",
	pages = "1--7",
	year = "2008"
}

@article{Colleaux:2015yta,
	author = "Coll{\'e}aux, Aimeric and Zerbini, sergio",
	title = "{Modified Gravity Models Admitting Second Order Equations of Motion}",
	eprint = "1508.06178",
	archivePrefix = "arXiv",
	primaryClass = "gr-qc",
	doi = "10.3390/e17106643",
	journal = "Entropy",
	volume = "17",
	number = "10",
	pages = "6643--6662",
	year = "2015"
}

@article{Gao:2012fd,
	author = "Gao, Changjun",
	title = "{Generalized modified gravity with the second order acceleration equation}",
	eprint = "1208.2790",
	archivePrefix = "arXiv",
	primaryClass = "gr-qc",
	doi = "10.1103/PhysRevD.86.103512",
	journal = "Phys. Rev. D",
	volume = "86",
	pages = "103512",
	year = "2012"
}

@article{Kobayashi:2011nu,
	author = "Kobayashi, Tsutomu and Yamaguchi, Masahide and Yokoyama, Jun'ichi",
	title = "{Generalized G-inflation: Inflation with the most general second-order field equations}",
	eprint = "1105.5723",
	archivePrefix = "arXiv",
	primaryClass = "hep-th",
	reportNumber = "KUNS-2339, RESCEU-9-11",
	doi = "10.1143/PTP.126.511",
	journal = "Prog. Theor. Phys.",
	volume = "126",
	pages = "511--529",
	year = "2011"
}

@phdthesis{Colleaux:2019ckh,
	author = "Colleaux, Aimeric",
	title = "{Regular black hole and cosmological spacetimes in Non-Polynomial Gravity theories}",
	school = "Trento U.",
	month = "6",
	year = "2019"
}

@article{Bueno:2025tli,
    author = "Bueno, Pablo and Hennigar, Robie A. and Murcia, {\'A}ngel J. and Vicente-Cano, Aitor",
    title = "{Buchdahl limits in theories with regular black holes}",
    eprint = "2512.19796",
    archivePrefix = "arXiv",
    primaryClass = "gr-qc",
    doi = "10.1103/mcrq-6fl3",
    journal = "Phys. Rev. D",
    volume = "113",
    number = "8",
    pages = "084008",
    year = "2026"
}

@article{Tan:2025hht,
	author = "Tan, Chen and Wang, Yong-Qiang",
	title = "{Frozen Neutron Stars in Four-Dimensional Non-polynomial Gravities}",
	eprint = "2512.23525",
	archivePrefix = "arXiv",
	primaryClass = "gr-qc",
	month = "12",
	year = "2025"
}

@article{Hawking:1970zqf,
	author = "Hawking, S. W. and Penrose, R.",
	title = "{The Singularities of gravitational collapse and cosmology}",
	doi = "10.1098/rspa.1970.0021",
	journal = "Proc. Roy. Soc. Lond. A",
	volume = "314",
	pages = "529--548",
	year = "1970"
}

@article{Carballo-Rubio:2025fnc,
	author = "Carballo-Rubio, Ra{\'u}l and others",
	title = "{Towards a non-singular paradigm of black hole physics}",
	eprint = "2501.05505",
	archivePrefix = "arXiv",
	primaryClass = "gr-qc",
	doi = "10.1088/1475-7516/2025/05/003",
	journal = "JCAP",
	volume = "05",
	pages = "003",
	year = "2025"
}

@inproceedings{Buoninfante:2024oxl,
	author = "Afshordi, Niayesh and others",
	editor = "Buoninfante, Luca and Carballo-Rubio, Ra{\'u}l and Cardoso, Vitor and Di Filippo, Francesco and Eichhorn, Astrid",
	title = "{Black Holes Inside and Out 2024: visions for the future of black hole physics}",
	eprint = "2410.14414",
	archivePrefix = "arXiv",
	primaryClass = "gr-qc",
	month = "10",
	year = "2024"
}

@inproceedings{Bambi:2025wjx,
	author = "Bambi, Cosimo and others",
	title = "{Black hole mimickers: from theory to observation}",
	eprint = "2505.09014",
	archivePrefix = "arXiv",
	primaryClass = "gr-qc",
	month = "5",
	year = "2025"
}

@inbook{Platania:2023srt,
	author = "Platania, Alessia",
	title = "{Black Holes in Asymptotically Safe Gravity}",
	eprint = "2302.04272",
	archivePrefix = "arXiv",
	primaryClass = "gr-qc",
	reportNumber = "NORDITA 2022-085",
	doi = "10.1007/978-981-19-3079-9_24-1",
	year = "2023"
}

@inbook{Buoninfante:2022ild,
	author = "Buoninfante, Luca and Giacchini, Breno L. and de Paula Netto, Tib{\'e}rio",
	title = "{Black Holes in Non-local Gravity}",
	eprint = "2211.03497",
	archivePrefix = "arXiv",
	primaryClass = "gr-qc",
	reportNumber = "NORDITA 2022-076",
	doi = "10.1007/978-981-19-3079-9_36-1",
	year = "2024"
}

@article{Lan:2023cvz,
	author = "Lan, Chen and Yang, Hao and Guo, Yang and Miao, Yan-Gang",
	title = "{Regular Black Holes: A Short Topic Review}",
	eprint = "2303.11696",
	archivePrefix = "arXiv",
	primaryClass = "gr-qc",
	doi = "10.1007/s10773-023-05454-1",
	journal = "Int. J. Theor. Phys.",
	volume = "62",
	number = "9",
	pages = "202",
	year = "2023"
}

@article{Bronnikov:2022ofk,
	author = "Bronnikov, Kirill A.",
	title = "{Regular black holes sourced by nonlinear electrodynamics}",
	eprint = "2211.00743",
	archivePrefix = "arXiv",
	primaryClass = "gr-qc",
	month = "11",
	year = "2022"
}

@article{Balart:2014cga,
	author = "Balart, Leonardo and Vagenas, Elias C.",
	title = "{Regular black holes with a nonlinear electrodynamics source}",
	eprint = "1408.0306",
	archivePrefix = "arXiv",
	primaryClass = "gr-qc",
	doi = "10.1103/PhysRevD.90.124045",
	journal = "Phys. Rev. D",
	volume = "90",
	number = "12",
	pages = "124045",
	year = "2014"
}

@article{Ayon-Beato:1998hmi,
	author = "Ayon-Beato, Eloy and Garcia, Alberto",
	title = "{Regular black hole in general relativity coupled to nonlinear electrodynamics}",
	eprint = "gr-qc/9911046",
	archivePrefix = "arXiv",
	doi = "10.1103/PhysRevLett.80.5056",
	journal = "Phys. Rev. Lett.",
	volume = "80",
	pages = "5056--5059",
	year = "1998"
}

@article{Ayon-Beato:1999qin,
	author = "Ayon-Beato, Eloy and Garcia, Alberto",
	title = "{Nonsingular charged black hole solution for nonlinear source}",
	eprint = "gr-qc/9911084",
	archivePrefix = "arXiv",
	doi = "10.1023/A:1026640911319",
	journal = "Gen. Rel. Grav.",
	volume = "31",
	pages = "629--633",
	year = "1999"
}

@article{Ayon-Beato:1999kuh,
	author = "Ayon-Beato, Eloy and Garcia, Alberto",
	title = "{New regular black hole solution from nonlinear electrodynamics}",
	eprint = "hep-th/9911174",
	archivePrefix = "arXiv",
	doi = "10.1016/S0370-2693(99)01038-2",
	journal = "Phys. Lett. B",
	volume = "464",
	pages = "25",
	year = "1999"
}

@article{Rodrigues:2018bdc,
	author = "Rodrigues, Manuel E. and de Sousa Silva, Marcos V.",
	title = "{Bardeen Regular Black Hole With an Electric Source}",
	eprint = "1802.05095",
	archivePrefix = "arXiv",
	primaryClass = "gr-qc",
	doi = "10.1088/1475-7516/2018/06/025",
	journal = "JCAP",
	volume = "06",
	pages = "025",
	year = "2018"
}

@article{Bronnikov:2000vy,
	author = "Bronnikov, Kirill A.",
	title = "{Regular magnetic black holes and monopoles from nonlinear electrodynamics}",
	eprint = "gr-qc/0006014",
	archivePrefix = "arXiv",
	doi = "10.1103/PhysRevD.63.044005",
	journal = "Phys. Rev. D",
	volume = "63",
	pages = "044005",
	year = "2001"
}

@article{Dymnikova:2004zc,
	author = "Dymnikova, Irina",
	title = "{Regular electrically charged structures in nonlinear electrodynamics coupled to general relativity}",
	eprint = "gr-qc/0407072",
	archivePrefix = "arXiv",
	doi = "10.1088/0264-9381/21/18/009",
	journal = "Class. Quant. Grav.",
	volume = "21",
	pages = "4417--4429",
	year = "2004"
}

@article{Zaslavskii:2010qz,
	author = "Zaslavskii, O. B.",
	title = "{Regular black holes and energy conditions}",
	eprint = "1004.2362",
	archivePrefix = "arXiv",
	primaryClass = "gr-qc",
	doi = "10.1016/j.physletb.2010.04.031",
	journal = "Phys. Lett. B",
	volume = "688",
	pages = "278--280",
	year = "2010"
}

@article{Balart:2023odm,
	author = "Balart, Leonardo and Panotopoulos, Grigoris and Rinc{\'o}n, {\'A}ngel",
	title = "{Regular Charged Black Holes, Energy Conditions, and Quasinormal Modes}",
	eprint = "2309.01910",
	archivePrefix = "arXiv",
	primaryClass = "gr-qc",
	doi = "10.1002/prop.202300075",
	journal = "Fortsch. Phys.",
	volume = "71",
	number = "12",
	pages = "2300075",
	year = "2023"
}

@article{Borissova:2025msp,
	author = "Borissova, Johanna and Liberati, Stefano and Visser, Matt",
	title = "{Violations of the null convergence condition in kinematical transitions between singular and regular black holes, horizonless compact objects, and bounces}",
	eprint = "2502.00548",
	archivePrefix = "arXiv",
	primaryClass = "gr-qc",
	doi = "10.1103/PhysRevD.111.104054",
	journal = "Phys. Rev. D",
	volume = "111",
	number = "10",
	pages = "104054",
	year = "2025"
}

@article{Frolov:2024hhe,
	author = "Frolov, Valeri P. and Koek, Alex and Soto, Jose Pinedo and Zelnikov, Andrei",
	title = "{Regular black holes inspired by quasitopological gravity}",
	eprint = "2411.16050",
	archivePrefix = "arXiv",
	primaryClass = "gr-qc",
	reportNumber = "Alberta Thy 11-24",
	doi = "10.1103/PhysRevD.111.044034",
	journal = "Phys. Rev. D",
	volume = "111",
	number = "4",
	pages = "044034",
	year = "2025"
}

@article{Giesel:2025kdl,
	author = "Giesel, Kristina and Liu, Hongguang",
	title = "{From Principles to Effective Models: A Constructive Framework for Effective Covariant Actions with a Unique Vacuum Solution}",
	eprint = "2512.24960",
	archivePrefix = "arXiv",
	primaryClass = "gr-qc",
	month = "12",
	year = "2025"
}

@article{Frolov:2026rcm,
    author = "Frolov, Valeri P. and Zelnikov, Andrei",
    title = "{Regular black holes in quasitopological gravity: Null shells and mass inflation}",
    eprint = "2601.01861",
    archivePrefix = "arXiv",
    primaryClass = "gr-qc",
    reportNumber = "Alberta Thy 5-25",
    doi = "10.1103/nrd7-j6hs",
    journal = "Phys. Rev. D",
    volume = "113",
    number = "8",
    pages = "084007",
    year = "2026"
}

@article{Markov:1982rcm,
	author = "Markov, M.A",
	title = "{Limiting density of matter as a universal law of nature}",
	journal = "Letters to ZhETF",
volume = "36",
pages = "214",
	year = "1982"
}

\end{document}